\documentclass[lettersize,journal]{IEEEtran}
\usepackage{amsmath,amsfonts,amssymb}
\usepackage{algorithmic}
\usepackage{algorithm}
\usepackage[dvipsnames]{xcolor}
\usepackage{array}
\usepackage{textcomp}
\usepackage{stfloats}
\usepackage{url}
\usepackage{verbatim}
\usepackage{graphicx}
\usepackage{cite}
\usepackage[para,online,flushleft]{threeparttable}
\usepackage{multirow}
\usepackage{booktabs}
\usepackage{tabularx,ragged2e,siunitx}
\usepackage{subcaption}
\usepackage{arydshln}
\newcommand{\dataStats}{
\begin{table}[!hbp]
  \centering
  \captionsetup[subfloat]{labelfont=scriptsize,textfont=scriptsize,justification=centering}
  \caption{Summary Statistics of the Selected Datasets}
  \begin{threeparttable}
    \begin{tabular}{p{1.6cm}p{0.7cm}p{0.9cm}p{0.6cm}p{0.9cm}p{0.5cm}p{0.5cm}}
    \toprule
    \multicolumn{2}{c}{Dataset} &  \#NTG   & \#Texts & ASL\tnote{1} & AScS\tnote{2} & AWS\tnote{3}\\
    \midrule
    AuTexT \cite{sarvazyan2023overview} & ~ & 6 & 55709 & 4670  & 4  & 64 \\
    SynSci \cite{rosati2022synscipass}& ~ & 5 & 5889  & 978   & 6    & 180 \\
    TweepF \cite{fagni2021tweepfake} & ~ & 3 & 25572 & 4262  & 2     & 23 \\
    Uchendu \cite{uchendu2020authorship}& ~ & 8 & 9594  & 1066  & 22  & 535 \\
    TBench \cite{uchendu2021turingbench}& ~ & 19 & 168612 & 8408  & 8  & 216 \\
    \multirow{10}{*}{DFTD\tnote{5} \cite{li2023mage}} 
    & CMV &  \multirow{10}{*}{27}  & 21547 & 559   & 16    & 314 \\
    & ELI5 &  & 34707 & 659   & 14   & 297 \\
    & SciGen & & 25133 & 678   & 10  & 257 \\
    & SQuAD &  & 36420 & 715   & 10  & 252 \\
    & TLDR &   & 19481 & 569   & 9   & 199 \\
    & WP  &  & 27731 & 678   & 35    & 504 \\
    & XSum &  & 27746 & 743   & 11    & 266 \\
    & Yelp &  & 43484 & 659   & 12   & 181 \\
    \multirow{3}{*} {FLAME} 
    & Pure & 25 & 59831 & 1343  & 20  & 350 \\
    & Perturb & 25(135)\tnote{4} & 83499 & 666(124)  & 22  & 394 \\
    & Human & 97\tnote{4} & 12010 & 124 & 19 & 342 \\
    \bottomrule
    \end{tabular}%
    \begin{tablenotes}
        \item[1] Average Samples per NTG; \item[2] Average Sentences per Sample;\item[3] Average Words per Sample; \item[4] Number of Human authors; \item[5] Dataset was pre-processed. Details described in Supplementary Materials     
    \end{tablenotes}
    \end{threeparttable}
  \label{tab:data_stats}%
\end{table}%
}

\newcommand{\detectorTable}{
\begin{table*}
\centering
    \captionsetup[subfloat]{labelfont=scriptsize,textfont=scriptsize,justification=centering}
    \caption{Brief description of the selected detectors}
    \begin{threeparttable}
    \begin{tabular}{p{0.04\textwidth}p{0.08\textwidth}p{0.81\textwidth}}
    \toprule
    Type                 & Det. Method     & Description \\ 
    \midrule
    \multirow{7}{*}{Metric}  
    & LL\tnote{1}  &  Average log probabilities of each word in text using GPT2-medium trained with Logistic Regression \cite{he2023mgtbench}. \\
    & Rank            &  Average absolute rank of each word based on preceding context using GPT2-medium trained with Logistic Regression \cite{he2023mgtbench}.  \\
    & Log Rank &  Average of logarithmic operator applied to the rank of each word using GPT2-medium trained with Logistic Regression \cite{he2023mgtbench}. \\
    & Entropy  &  Average entropy of each word conditioned on preceding context using GPT2-medium trained with Logistic Regression \cite{he2023mgtbench}.\\
    & GLTR  &  Test2 features from \cite{gehrmann2019gltr} using GPT2-large trained with grid search hyperparameter tuned Logistic Regression \cite{pu2023deepfake}.           \\
    & DetectGPT   & Log probabilities of words using GPT2-medium from random text perturbations using T5. Decisions made via thresholding \cite{mitchell2023detectgpt} \\
    & DetectLLM\tnote{2}   & Ratio of log likelihood and log rank (LRR) using GPT2-XL. Decision employs thresholding using AUROC curve \cite{su2023detectllm}            \\ \hline\noalign{\vskip 0.5ex}
    \multirow{4}{*}{Content} 
    & LIWC-DT & LIWC psycho-linguistic features \cite{boyd2022development} trained with Decision Trees classifier. \\                   
    & WRFC\tnote{3}  &  Writeprints stylometric features \cite{abbasi2008writeprints} trained with Random Forest Classifier \cite{uchendu2020authorship}. \\
    & LFI  &  21 dimensional feature set comprising of stylistic, complexity and psychological features trained on 3-layer CNN classifier \cite{aich2022demystifying}.        \\ 
    & Char-3gram\tnote{4} &   SVM classifier trained on character 3-gram features using one-vs-rest classification strategy.\cite{kestemont2019overview}  \\\hline\noalign{\vskip 0.5ex}
    \multirow{9}{*}{Model}   
    & AIGC   &  Multi-scale positive-unlabeled (MPU) framework applied on RoBERTa model for NTD \cite{tian2023multiscale}.  \\
    & ContraX  &  Contrastive learning with DeBERTa for cross-entropy fine-tuning, followed by classification using a 2-layer MLP. \cite{ai2022whodunit}        \\
    & BertAA  &  Cascaded architecture with BERT-FT classifier, stylistic and hybrid features trained using Logistic Regression. \cite{fabien2020bertaa}           \\
    & FFLM & Fine-tuned RoBERTa embeddings trained on a 2-layer stacked CNN classifier. \cite{munir2021through}\\
    & T5S\tnote{5}   & Fine-tuning T5 model to perform next-token-prediction for NTD and AA tasks\cite{chen2023token}.  \\
    & DeB-FT\tnote{6}  & Fine-tuned DeBERTa model with a multi-class classification layer on top of it.\cite{rosati2022synscipass}            \\
    & XLNet-FT & Fine-tuned XLNet model with a multi-class classification layer on top of it. \cite{uchendu2021turingbench}\\ 
    & BERT-FT  & \multirow{2}{*}{Fine-tuned BERT model with a binary \cite{ippolito2020automatic} and multi-class \cite{uchendu2021turingbench} classification layer on top}            \\
    & BERT-multi\\
    & openAI-D\tnote{7} & \multirow{2}{*}{Fine-tuned RoBERTa model with a binary \cite{solaiman2019release} and multi-class \cite{uchendu2021turingbench} classification layer on top. }\\
    & RoB-multi\tnote{8}\\ 
    
    \bottomrule
    \end{tabular}
    \begin{tablenotes}
    \item[1]Log Likelihood; \item[2]detectLLM-LRR; \item[3]WriteprintsRFC; 
    \item[4]Character-3-grams; \item[5]T5-sentinel;  \item[6]DeBERTa-FineTuned;  
    \item[7]openAI-Detector; \item[8]RoBERTa-multi
   
    \end{tablenotes}
    \end{threeparttable}
    \label{tab:detectors}
\end{table*}
}

\newcommand{\IdealResults}{
\begin{figure}[]
\centering
\captionsetup[subfloat]{labelfont=scriptsize,textfont=scriptsize,justification=centering}

\subfloat[Incident Detection] {{\label{fig:ntd_ideal}} \includegraphics[clip,width=0.9\columnwidth]{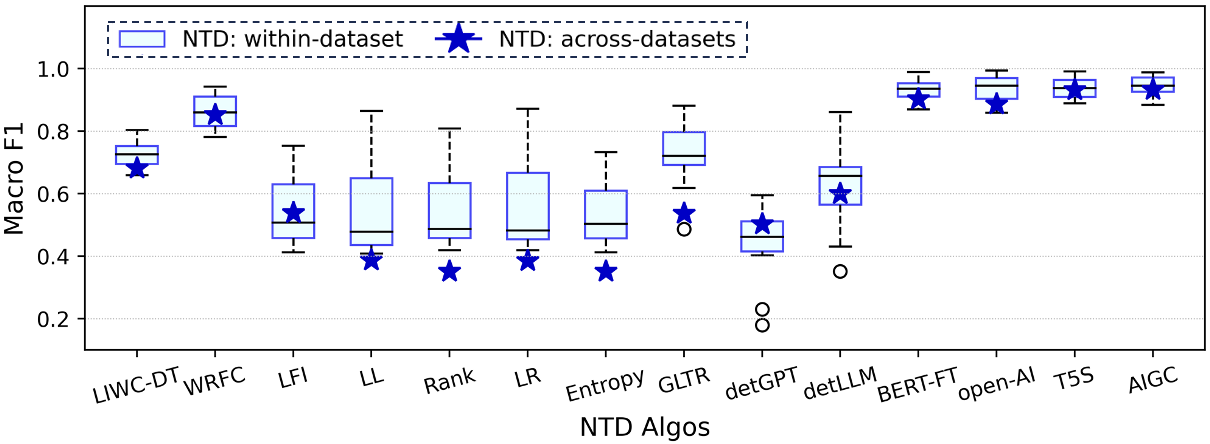}}

\subfloat[Forensic Profiling] {{\label{fig:aa_ideal}} \includegraphics[clip,width=0.9\columnwidth]{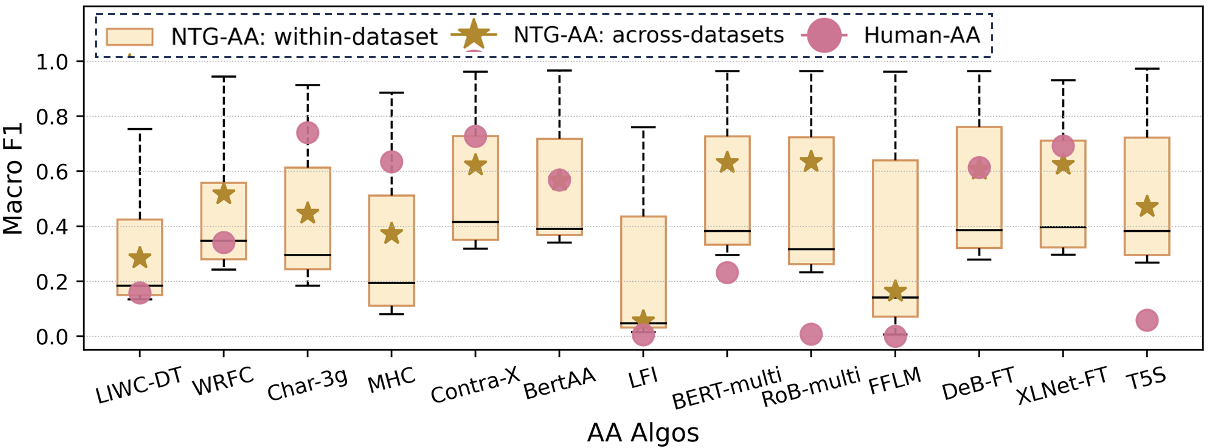}}

\caption{Results under Ideal Scenarios reported in Macro-F1 score}
\label{fig:ideal_results}
\end{figure}
}

\newcommand{\AttackResults}{
\begin{figure}
\centering
\includegraphics[width=0.8\linewidth]{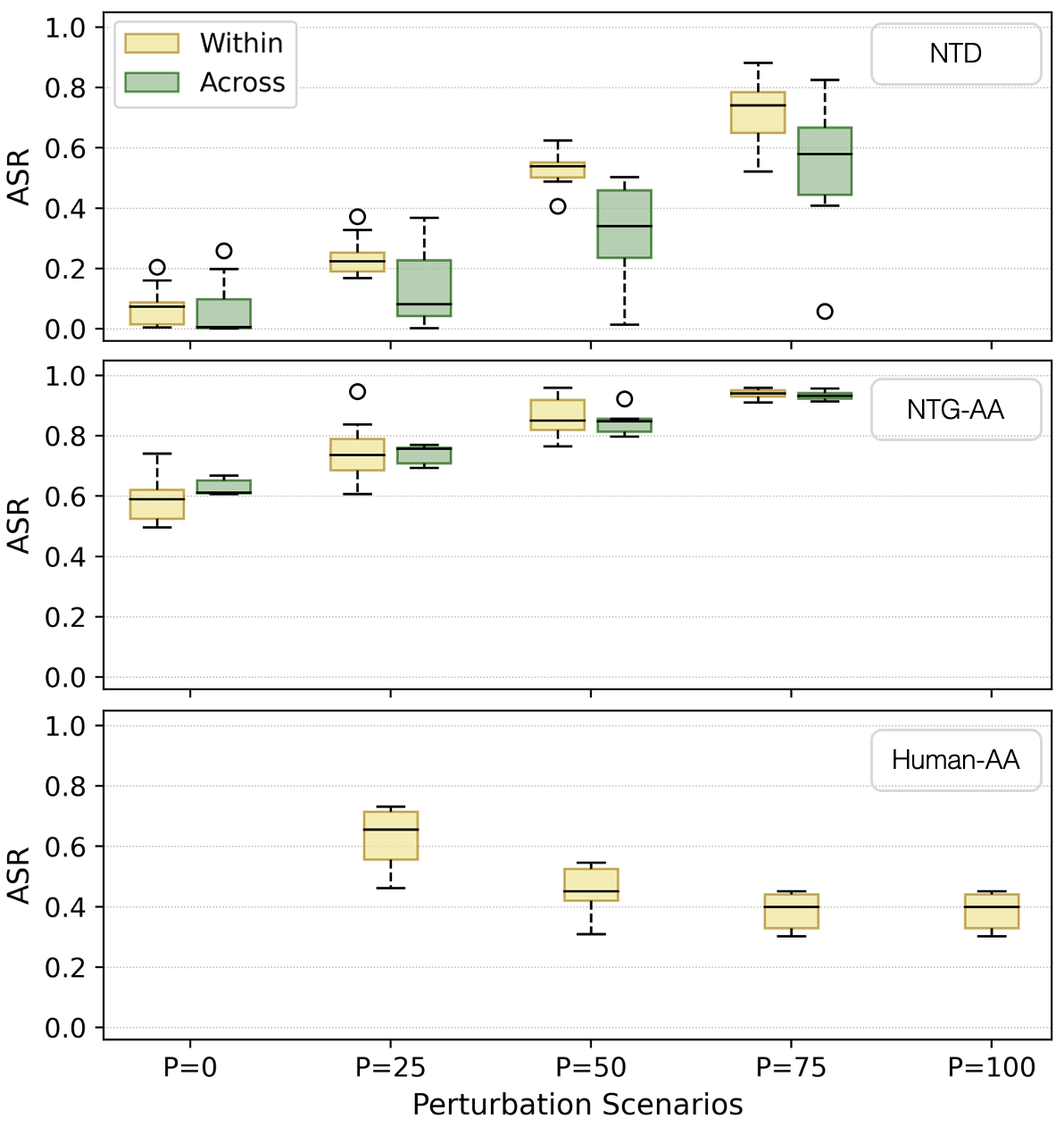}
\caption{Results under CS-ACT Attack Scenarios. ``P'' represents the proportion of human in the human-NTG co-authorship.}
\label{fig:attack_results}
\end{figure}
}

\newcommand{\FLAMEOrg}{
\begin{figure}
    \centering
    \begin{threeparttable}
        \includegraphics[width=0.8\linewidth]{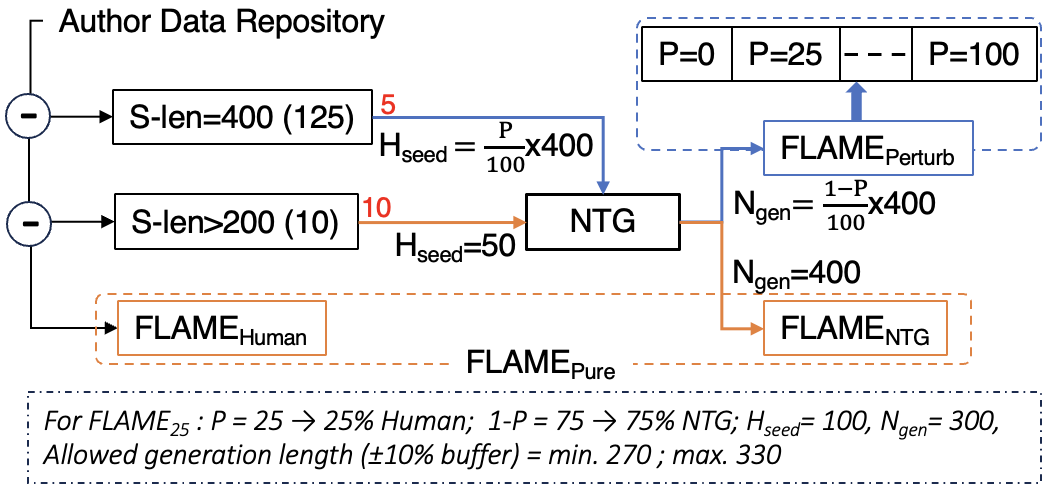}
        
        \begin{tablenotes}
            \footnotesize
            \item[] Note: Number in ``( )'' reflects the total number of author samples randomly selected for the task and numbers in \textcolor{red}{red} are samples given to each NTG for generation.
            \item[] Acronyms: S-len - Sample length requirement; 
            \item[] H$_{seed}$ - Human seed text length; 
            \item[] N$_{gen}$ - NTG generated text length   
        \end{tablenotes}
    \end{threeparttable}
\caption{Organization and Creation Process for FLAME dataset}
\label{fig:flame}
\end{figure}

}

\newcommand{\DFIR}{
\begin{figure*}
    \centering
    \includegraphics[width=0.8\textwidth]{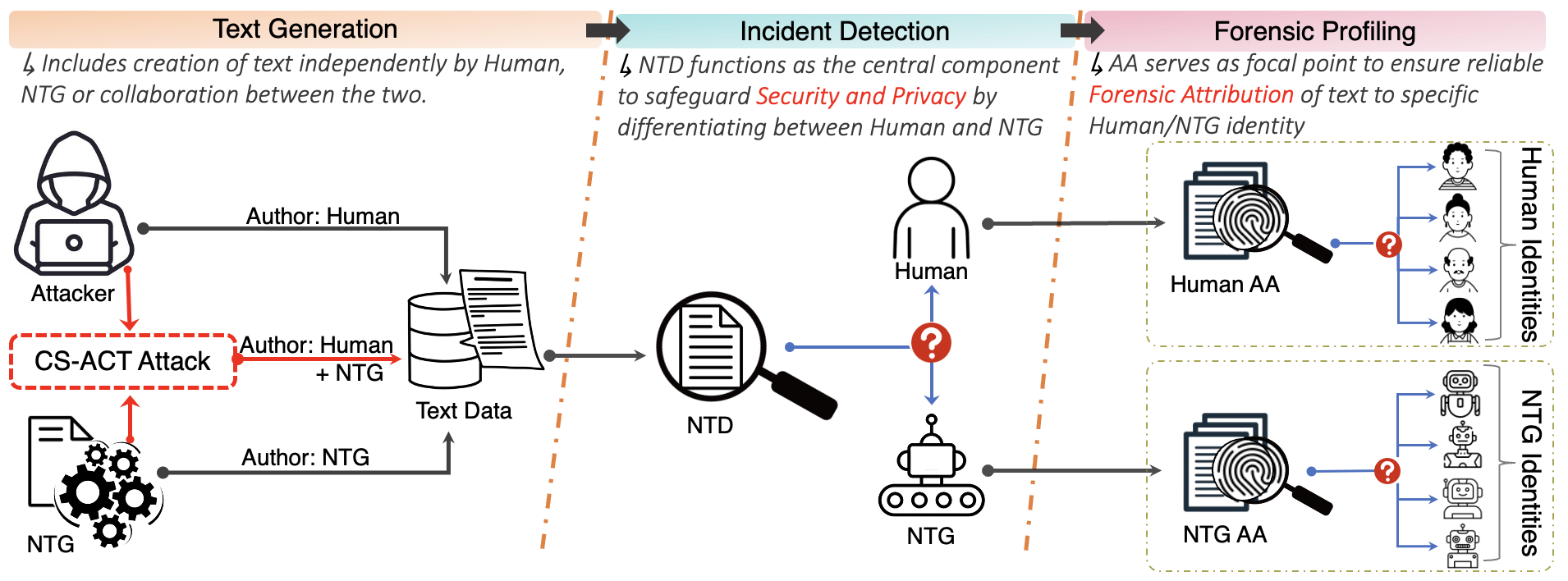}
    \caption{Overview of the DFIR pipeline}
    \label{fig:dfir}
\end{figure*}
}

\newcommand{\ConfusionMatrix}{
\begin{figure}[]
\centering
\captionsetup[subfloat]{labelfont=scriptsize,textfont=scriptsize,justification=centering}

\subfloat[GPT3 models] {{\label{fig:gpt3_modelfam}} \includegraphics[clip,width=0.48\columnwidth]{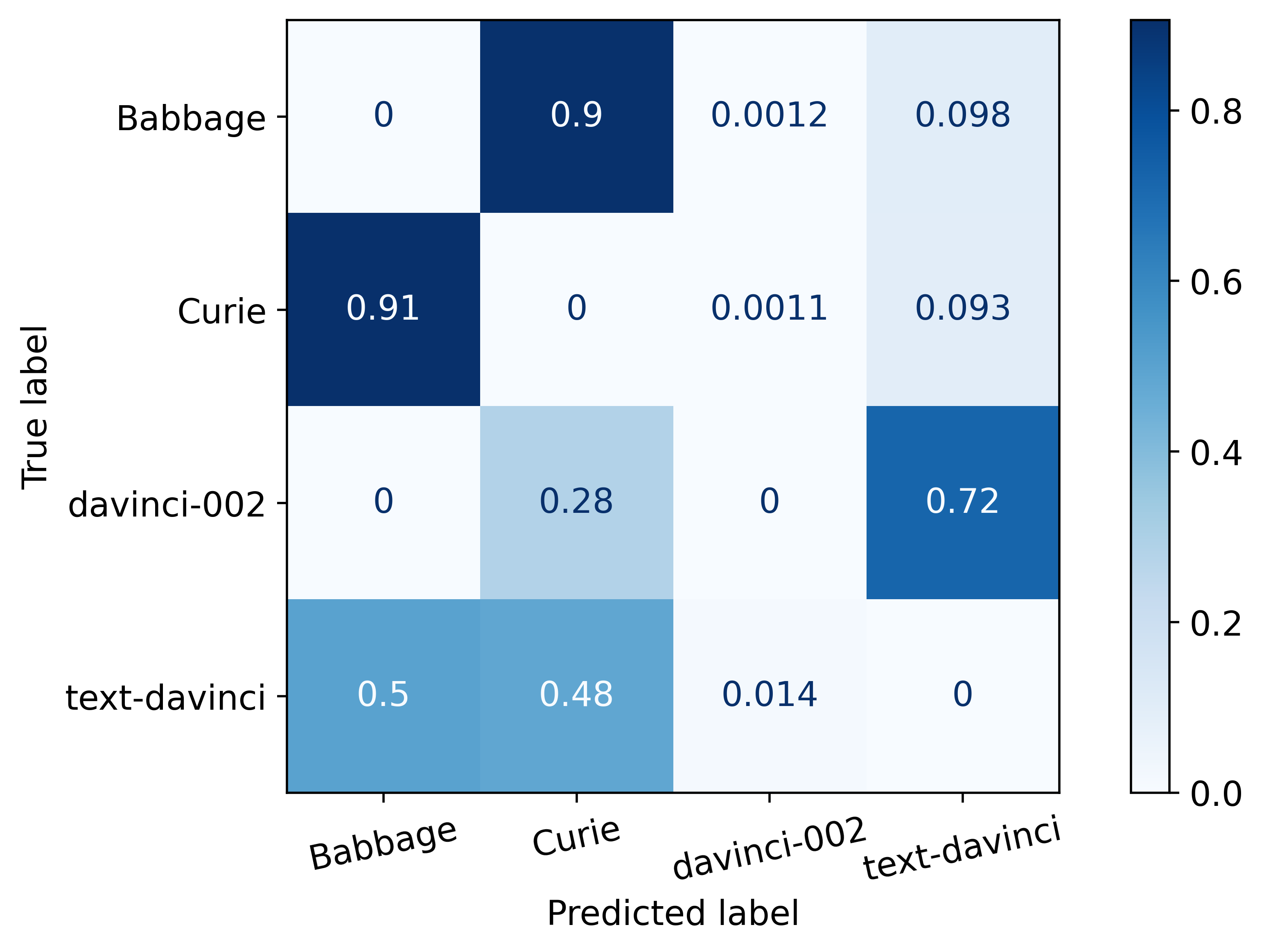}}
\hfill
\subfloat[Llama2-FT models] {{\label{fig:llama_model_fam}} \includegraphics[clip,width=0.48\columnwidth]{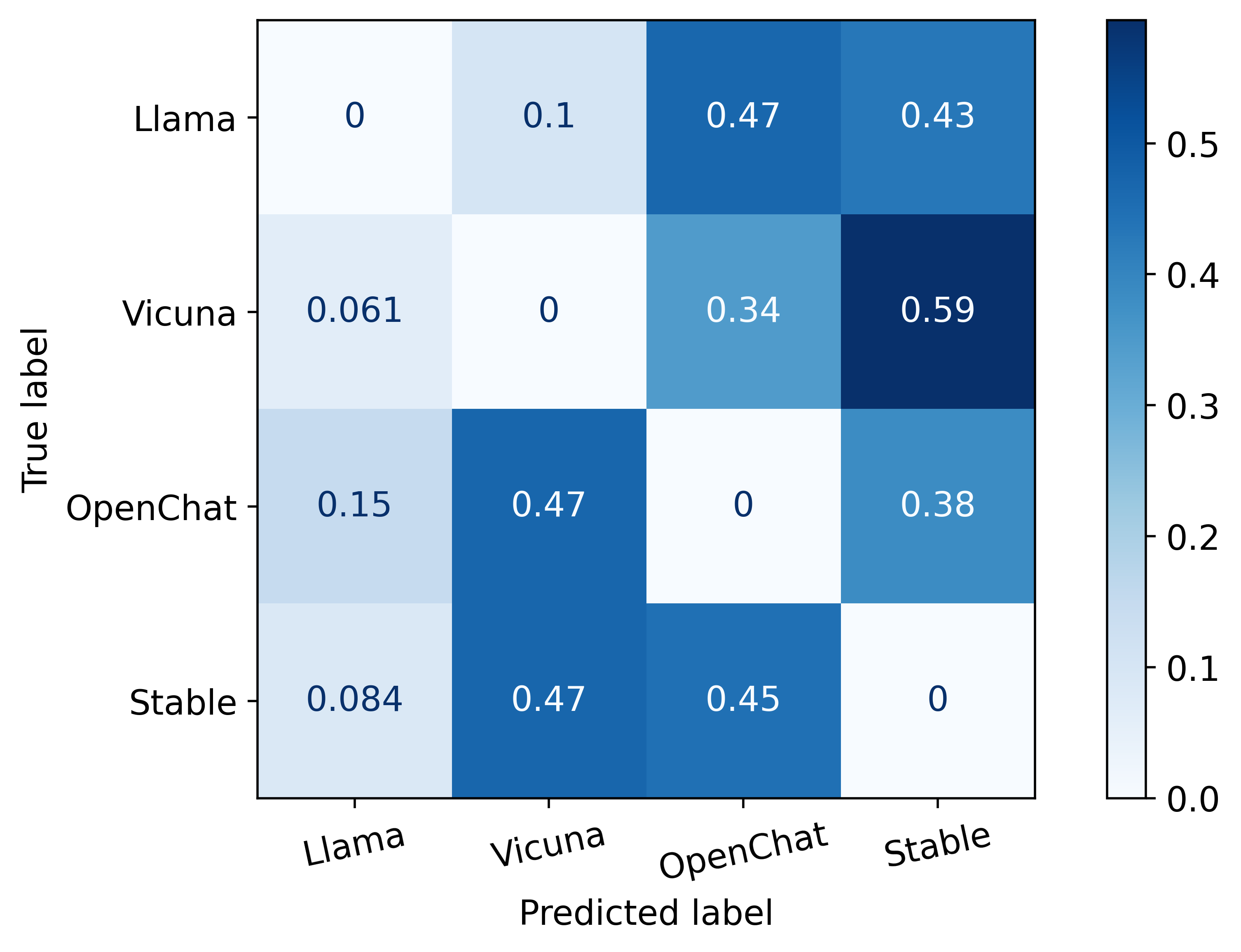}}

\caption{Confusion Matrix of Misclassifications within NTG family }
\label{fig:conf_mat}
\end{figure}
}

\newcommand{\LLMwiseNTGAA}{
\begin{figure*}[ht!]
    \centering
    \includegraphics[width=0.95\textwidth]{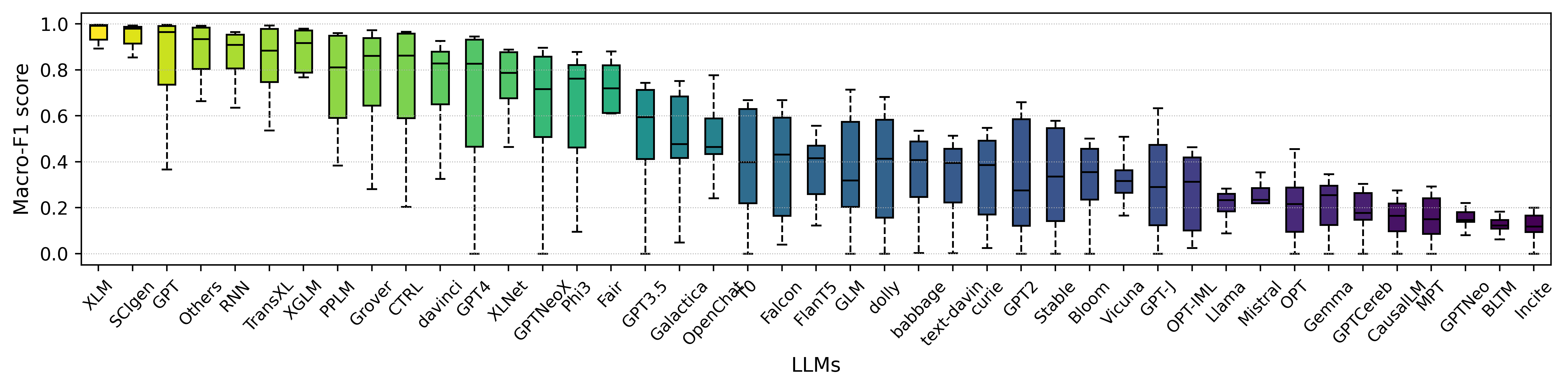}
    \caption{Performance for each LLM in Across-datasets scenario for NTG-AA}
    \label{fig:LLMwiseNTGAA_across}
\end{figure*}
}

\newcommand{\EffectModelVariant}{
\begin{table}[ht]
    \centering
    \caption{Effect of Diverse NTG Variants}
    \begin{threeparttable}
    \begin{tabular}{p{1.5cm}p{1.5cm}p{1.5cm}p{1cm}p{1cm}}
    \toprule
    Dataset   & Base Model & Spearman-R\tnote{1} & $\text{Mis}_{\text{T}}$\tnote{2} & $\text{Mis}_{\text{V}}$\tnote{3}  \\
    \midrule
    \multirow{3}{*}{DFTD}   & OPT  & \textbf{-0.22**}  &  0.72     & 0.64    \\
                         & Llama-2 & \textbf{-0.25**}   &   0.69    & 0.53    \\
                        & T5        & \textbf{0.33**}  &   0.78    &  0.36   \\
    AuText                  & Bloom  & \textbf{-0.41**}    &  0.64     &  0.76  \\
    \multirow{3}{*}{TBench} & Grover  & -0.13   &  0.72     &  0.89  \\
                        & GPT2  & -0.04    &   0.72    &  0.76 \\
                        & XLNet  & \textbf{0.47*}    &  0.19     &  0.34  \\ 
    \bottomrule
    \end{tabular}
    \begin{tablenotes}
    
    \item[1] Correlation between NTG Size and Performance; 
    for each base NTG: \item[2]proportion of total misclassifications and \item[3]proportion of misclassifications between its variants.
    \item[] Note: Values in \textbf{bold} reflect statistically significant correlations. \textbf{*} and \textbf{**} indicates significance at p$<$0.05 and p$<$0.01 respectively.
    \end{tablenotes}
    \end{threeparttable}
    \label{tab:effect_model_variant}
\end{table}
}

\newcommand{\EffectTextQuality}{
\begin{table}[ht]
    \centering
    \caption{Relationship between Text Quality and NTG-AA}
    \begin{threeparttable}
    \begin{tabular}{p{3.5cm}p{1.9cm}p{1.9cm}}
    \toprule
    Metric & Spearman-$R$\tnote{1} & Spearman-$\text{R}_{Z}$\tnote{1} \\
    \midrule
    Word Coherence & \textbf{-0.61} & \textbf{0.73}\\
    Sentence Coherence & \textbf{-0.61} & \textbf{0.71} \\
    Grammaticality & \textbf{-0.57} & \textbf{0.69}\\
    Redundancy & -0.26 & 0.09\\
    SQSE & 0.20 & \textbf{0.59}\\    
    \bottomrule
    \end{tabular}
    \begin{tablenotes}
    \item[1] Correlation between text quality metric and NTG macro-F1;
    \item[2] Correlation between Z-score and NTG macro-F1. Z-scores are calculated as $z'=\text{min}(\mid(\text{x}-\mu)\sigma^{-1}\mid,\text{3})$ where $\mu$ and $\sigma$ are the mean and standard deviation of the human distribution.
    \item Note: Values in \textbf{bold} font indicate statistically significant (p$\leq$0.01) correlations.
    \end{tablenotes}
    \end{threeparttable}
    \label{tab:text_quality}
\end{table}
}

\newcommand{\detectASR}{
\begin{table*}[htp!]
\centering
\captionsetup[subfloat]{labelfont=scriptsize,textfont=scriptsize,justification=centering}
\caption{NTD Performance Results for Ideal Scenario Measured in Macro-F1}
\begin{threeparttable}[t]
\begin{tabular}{lcccccccccccccc|c}

\toprule
Det. Methods & LIWC & WRFC\tnote{1} & LFI & LL\tnote{2} & Rank & LR\tnote{3} & Entropy & GLTR & detect & detect & BERT- & open- & T5S\tnote{7} & AIGC &  Avg./\\
Datasets ($\downarrow$)  & DT   &   &   &   &  &    &   &  & GPT & LLM\tnote{6}  & FT\tnote{4} & AI\tnote{5} &     &   &dataset\\
\midrule
AuText & 0.66 & 0.78 & 0.60 & 0.66 & 0.59 & 0.68 & 0.53 & 0.73 & 0.60 & 0.72 & 0.87 & 0.86 & 0.91 & 0.90 & 0.73 \\ 
SynSci & 0.72 & 0.94 & 0.45 & 0.46 & 0.46 & 0.46 & 0.45 & 0.62 & 0.44 & 0.63 & 0.97 & 0.99 & 0.99 & 0.98 & 0.68 \\
TFake & 0.72 & 0.84 & 0.64 & 0.69 & 0.64 & 0.67 & 0.67 & 0.69 & 0.41 & 0.43 & 0.89 & 0.90 & 0.89 & 0.88 & 0.68 \\ 
Uchendu & 0.74 & 0.92 & 0.47 & 0.47 & 0.47 & 0.47 & 0.47 & 0.71 & 0.23 & 0.43 & 0.93 & 0.97 & 0.94 & 0.94 & 0.65 \\
TBench & 0.76 & 0.84 & 0.49 & 0.49 & 0.49 & 0.49 & 0.49 & 0.49 & 0.18 & 0.35 & 0.99 & 0.99 & 0.99 & 0.99 & 0.64 \\ 
CMV & 0.78 & 0.89 & 0.41 & 0.41 & 0.44 & 0.42 & 0.41 & 0.83 & 0.44 & 0.63 & 0.95 & 0.96 & 0.96 & 0.97 & 0.70 \\ 
ELI5 & 0.74 & 0.88 & 0.64 & 0.59 & 0.66 & 0.61 & 0.58 & 0.80 & 0.52 & 0.64 & 0.95 & 0.95 & 0.95 & 0.96 & 0.77 \\
SciGen & 0.67 & 0.80 & 0.42 & 0.43 & 0.47 & 0.45 & 0.43 & 0.71 & 0.48 & 0.69 & 0.91 & 0.88 & 0.90 & 0.93 & 0.67 \\
SQuAD & 0.73 & 0.87 & 0.75 & 0.61 & 0.62 & 0.65 & 0.64 & 0.78 & 0.40 & 0.71 & 0.94 & 0.94 & 0.94 & 0.95 & 0.78 \\ 
TLDR & 0.70 & 0.81 & 0.44 & 0.45 & 0.49 & 0.48 & 0.52 & 0.64 & 0.49 & 0.68 & 0.92 & 0.92 & 0.92 & 0.93 & 0.68 \\ 
WP & 0.80 & 0.92 & 0.50 & 0.41 & 0.45 & 0.43 & 0.48 & 0.86 & 0.49 & 0.67 & 0.95 & 0.96 & 0.96 & 0.97 & 0.72 \\ 
XSum & 0.69 & 0.81 & 0.52 & 0.42 & 0.42 & 0.42 & 0.42 & 0.71 & 0.56 & 0.54 & 0.90 & 0.87 & 0.89 & 0.92 & 0.66 \\ 
Yelp & 0.69 & 0.85 & 0.51 & 0.67 & 0.70 & 0.68 & 0.62 & 0.75 & 0.54 & 0.67 & 0.92 & 0.91 & 0.94 & 0.94 & 0.76 \\
FLAME$_{Pure}$ & 0.78 & 0.93 & 0.71 & 0.87 & 0.81 & 0.87 & 0.73 & 0.88 & 0.44 & 0.86 & 0.98 & 0.99 & 0.99 & 0.99 & 0.87 \\ \hdashline\noalign{\vskip 0.5ex}
Avg./Method & 0.73 & 0.86 & 0.54 & 0.54 & 0.55 & 0.56 & 0.53 & 0.73 & 0.31 & 0.62 & 0.93 & 0.94 & 0.94 & 0.95 & ~ \\ \hline
Across\tnote{8} & 0.68 & 0.85 & 0.54 & 0.39 & 0.35 & 0.39 & 0.35 & 0.54 & 0.50 & 0.60 & 0.90 & 0.89 & 0.93 & 0.93 & 0.63 \\ 
\bottomrule
\end{tabular}%
\begin{tablenotes}
\item[1] WriteprintsRFC; \item[2] Log Likelihood; \item[3]Log Rank; \item[4] BERT-FineTuned; \item[5] openAI-Detector; \item[6] DetectLLM-LRR; \item[7] T5-sentinel; \item[8] Across-datasets 
\end{tablenotes}
\end{threeparttable}
\label{tab:ASR_detect}%
\end{table*}
}

\newcommand{\NTGAA}{
\begin{table*}[htp!]
\centering
\captionsetup[subfloat]{labelfont=scriptsize,textfont=scriptsize,justification=centering}
\caption{AA Performance Results for Ideal Scenario Measured in Macro-F1}
\begin{threeparttable}[t]
\begin{tabular}{lccccccccccccc|c}

\toprule
AA Methods & LIWC & WRFC\tnote{1} & Char-3 & MHC & ContraX & BERT- & LFI & BERT- & RoB- & FFLM &  DeB- & XLNet- & T5S\tnote{5} & Avg./\\
Datasets ($\downarrow$)     & DT   &      &   grams\tnote{2} &      &  &   AA   &  & multi & multi\tnote{3}  &  & FT\tnote{4}    &   FT  &      &  dataset\\
\midrule

AuText &0.29 & 0.47 & 0.39 & 0.42 & 0.53 & 0.59 & 0.35 & 0.57 & 0.51 & 0.56 & 0.56 & 0.51 & 0.55 & 0.48 \\ 
SynSci & 0.59 & 0.79 & 0.73 & 0.63 & 0.90 & 0.82 & 0.53 & 0.86 & 0.91 & 0.91 & 0.90 & 0.85 & 0.81 & 0.79 \\ 
TFake & 0.75 & 0.87 & 0.89 & 0.88 & 0.96 & 0.97 & 0.76 & 0.96 & 0.97 & 0.96 & 0.97 & 0.93 & 0.95 & 0.91 \\ 
Uchendu & 0.73 & 0.94 & 0.91 & 0.89 & 0.94 & 0.95 & 0.73 & 0.96 & 0.94 & 0.96 & 0.94 & 0.93 & 0.97 & 0.91 \\ 
TBench & 0.47 & 0.59 & 0.69 & 0.54 & 0.79 & 0.76 & 0.47 & 0.78 & 0.79 & 0.67 & 0.83 & 0.78 & 0.78 & 0.69 \\
CMV & 0.17 & 0.30 & 0.25 & 0.10 & 0.35 & 0.37 & 0.03 & 0.35 & 0.31 & 0.10 & 0.32 & 0.34 & 0.29 & 0.25 \\
ELI5 & 0.13 & 0.24 & 0.18 & 0.08 & 0.32 & 0.34 & 0.03 & 0.31 & 0.23 & 0.04 & 0.28 & 0.30 & 0.27 & 0.21 \\ 
SciGen & 0.19 & 0.37 & 0.33 & 0.22 & 0.46 & 0.46 & 0.04 & 0.33 & 0.26 & 0.03 & 0.44 & 0.46 & 0.42 & 0.31 \\ 
SQuAD & 0.17 & 0.32 & 0.25 & 0.17 & 0.35 & 0.39 & 0.05 & 0.38 & 0.30 & 0.07 & 0.32 & 0.31 & 0.32 & 0.26 \\
TLDR & 0.14 & 0.28 & 0.21 & 0.11 & 0.35 & 0.36 & 0.03 & 0.30 & 0.25 & 0.07 & 0.32 & 0.33 & 0.29 & 0.23 \\
WP & 0.19 & 0.34 & 0.26 & 0.14 & 0.37 & 0.38 & 0.03 & 0.39 & 0.32 & 0.16 & 0.33 & 0.33 & 0.35 & 0.27 \\
XSum & 0.14 & 0.26 & 0.20 & 0.11 & 0.35 & 0.37 & 0.02 & 0.36 & 0.25 & 0.13 & 0.33 & 0.31 & 0.30 & 0.24 \\
Yelp & 0.14 & 0.27 & 0.24 & 0.10 & 0.34 & 0.36 & 0.05 & 0.31 & 0.28 & 0.01 & 0.32 & 0.32 & 0.29 & 0.23 \\
FLAME-NTG & 0.18 & 0.36 & 0.34 & 0.33 & 0.49 & 0.40 & 0.05 & 0.48 & 0.53 & 0.43 & 0.53 & 0.51 & 0.49 & 0.39 \\ \hdashline\noalign{\vskip 0.5ex}
Avg./Method & 0.31 & 0.46 & 0.42 & 0.34 & 0.54 & 0.54 & 0.23 & 0.52 & 0.49 & 0.36 & 0.53 & 0.51 & 0.51 & ~ \\ \hline
Across\tnote{6} & 0.28 & 0.52 & 0.45 & 0.37 & 0.62 & 0.56 & 0.06 & 0.63 & 0.63 & 0.16 & 0.61 & 0.62 & 0.47 & 0.46 \\ \hline
FLAME$_{H}$\tnote{7} & 0.16 & 0.34 & 0.74 & 0.63 & 0.73 & 0.57 & 0.01 & 0.23 & 0.01 & 0.00 & 0.61 & 0.69 & 0.06 & 0.37 \\ 
\bottomrule
\end{tabular}%
\begin{tablenotes}
\item[1] WriteprintsRFC; \item[2]Character-3-grams ; \item[3] RoBERTa-multi ; \item[4] DeBERTa-Finetuned; \item[5] T5-sentinel; \item[6] Across-datasets; \item[7] FLAME$_{Human}$
\end{tablenotes}
\end{threeparttable}
\label{tab:NTGAA_ASR}%
\end{table*}
}

\newcommand{\detectAttack}{
\begin{table*}[htp!]
\centering
\captionsetup[subfloat]{labelfont=scriptsize,textfont=scriptsize,justification=centering}
\caption{NTD Performance Results under CS-ACT Attack reported in ASR}
\begin{threeparttable}[t]
\begin{tabular}{lcccccccccccccc|c}

\toprule
Det. Methods & LIWC & WRFC\tnote{2} & LFI & LL\tnote{3} & Rank & LR\tnote{4} & Entropy & GLTR & detect & detect & BERT- & open- & T5S\tnote{8} & AIGC &  Avg./\\
Datasets ($\downarrow$)  & DT   &   &   &   &  &    &   &  & GPT & LLM\tnote{7}  & FT\tnote{5} & AI\tnote{6} &     &   &dataset\\

\midrule

FLAME-0 & 0.20 & 0.08 & 0.16 & 0.07 & 0.08 & 0.07 & 0.13 & 0.07 & 0.31 & 0.09 & 0.00 & 0.00 & 0.00 & 0.02 & 0.09 \\
Across-0\tnote{1} & 0.26 & 0.10 & 0.20 & 0.00 & 0.00 & 0.00 & 0.00 & 0.00 & 0.31 & 0.04 & 0.00 & 0.00 & 0.00 & 0.01 & 0.07 \\ \hdashline\noalign{\vskip 0.5ex}

FLAME-25 & 0.37 & 0.25 & 0.28 & 0.19 & 0.25 & 0.18 & 0.33 & 0.19 & 0.19 & 0.24 & 0.19 & 0.18 & 0.17 & 0.22 & 0.23 \\
Across-25 & 0.37 & 0.23 & 0.29 & 0.00 & 0.00 & 0.00 & 0.00 & 0.00 & 0.19 & 0.08 & 0.04 & 0.03 & 0.06 & 0.11 & 0.10 \\\hdashline\noalign{\vskip 0.5ex}

FLAME-50 & 0.55 & 0.54 & 0.41 & 0.51 & 0.49 & 0.50 & 0.53 & 0.55 & 0.13 & 0.56 & 0.56 & 0.55 & 0.50 & 0.63 & 0.50 \\
Across-50 & 0.49 & 0.46 & 0.36 & 0.00 & 0.00 & 0.00 & 0.00 & 0.01 & 0.14 & 0.24 & 0.33 & 0.23 & 0.34 & 0.50 & 0.22 \\\hdashline\noalign{\vskip 0.5ex}

FLAME-75 & 0.65 & 0.72 & 0.52 & 0.74 & 0.65 & 0.74 & 0.63 & 0.77 & 0.13 & 0.77 & 0.84 & 0.80 & 0.79 & 0.88 & 0.69 \\
Across-75 & 0.58 & 0.64 & 0.41 & 0.00 & 0.00 & 0.00 & 0.00 & 0.06 & 0.14 & 0.44 & 0.67 & 0.58 & 0.69 & 0.83 & 0.36 \\

\bottomrule
\end{tabular}%
\begin{tablenotes}
\item[1] Number following `-' represents FLAME$_{Perturb}$ test case. For instance, Across-0 presents results for methods trained on Across super-set and tested on FLAME$_{0}$; \item[2] WriteprintsRFC; \item[3] Log Likelihood; \item[4] Log Rank; \item[5] BERT-FineTuned; \item[6] openAI-Detector; \item[7] DetectLLM-LRR; \item[8] T5-sentinel
\end{tablenotes}
\end{threeparttable}
\label{tab:detect_attack}%
\end{table*}
}

\newcommand{\AAAttack}{
\begin{table*}[htbp]
\centering
\captionsetup[subfloat]{labelfont=scriptsize,textfont=scriptsize,justification=centering}
\caption{AA Performance Results under CS-ACT Attack reported in ASR}
\begin{threeparttable}[t]

\begin{tabular}{lccccccccccccc|c}
\toprule
AA Methods & LIWC & WRFC\tnote{2} & Char-3 & MHC & ContraX & BERTAA & LFI & BERT- & RoB- & FFLM &  DeB- & XLNet- & T5S\tnote{5} & Avg./\\
Datasets ($\downarrow$)     & DT   &      &   grams &      &  &      & &multi & multi\tnote{3}  &  & FT\tnote{4}    &   FT  &      & dataset\\
\midrule

& \multicolumn{13}{c}{NTG-AA} &\\ \hline\noalign{\vskip 0.5ex}

FLAME-0 & 0.84 & 0.74 & 0.66 & 0.67 & 0.53 & 0.65 & 0.92 & 0.59 & 0.59 & 0.67 & 0.50 & 0.52 & 0.52 & 0.65 \\
Across-0 & 0.89 & 0.77 & 0.72 & 0.75 & 0.61 & 0.72 & 0.97 & 0.65 & 0.67 & 0.91 & 0.61 & 0.61 & 0.72 & 0.74 \\ \hdashline\noalign{\vskip 0.5ex}

FLAME-25\tnote{1} & 0.90 & 0.84 & 0.76 & 0.81 & 0.74 & 0.95 & 0.94 & 0.74 & 0.74 & 0.79 & 0.61 & 0.64 & 0.60 & 0.77 \\
Across-25 & 0.92 & 0.86 & 0.78 & 0.83 & 0.76 & 0.95 & 0.96 & 0.76 & 0.77 & 0.94 & 0.69 & 0.71 & 0.76 & 0.82 \\\hdashline\noalign{\vskip 0.5ex}

FLAME-50 & 0.94 & 0.91 & 0.84 & 0.91 & 0.93 & 0.96 & 0.95 & 0.85 & 0.84 & 0.87 & 0.77 & 0.80 & 0.72 & 0.87 \\
Across-50 & 0.95 & 0.91 & 0.85 & 0.91 & 0.92 & 0.96 & 0.96 & 0.86 & 0.85 & 0.95 & 0.80 & 0.81 & 0.83 & 0.89 \\\hdashline\noalign{\vskip 0.5ex}

FLAME-75 & 0.95 & 0.94 & 0.92 & 0.95 & 0.96 & 0.96 & 0.96 & 0.94 & 0.93 & 0.94 & 0.91 & 0.93 & 0.86 & 0.94 \\
Across-75 & 0.96 & 0.95 & 0.92 & 0.95 & 0.96 & 0.96 & 0.96 & 0.94 & 0.92 & 0.97 & 0.91 & 0.93 & 0.91 & 0.94 \\ \hline\noalign{\vskip 0.5ex}

& \multicolumn{13}{c}{Human-AA} & \\ \hline\noalign{\vskip 0.5ex}

FLAME-25 & 0.93 & 0.87 & 0.63 & 0.73 & 0.53 & 0.46 & 0.98 & 0.83 & 0.98 & 0.99 & 0.73 & 0.68 & 0.96 & 0.79  \\
FLAME-50 & 0.90 & 0.79 & 0.41 & 0.55 & 0.31 & 0.44 & 0.98 & 0.78 & 0.97 & 0.99 & 0.55 & 0.46 & 0.94 & 0.70   \\ 
FLAME-75 & 0.87 & 0.73 & 0.32 & 0.44 & 0.30 & 0.44 & 0.98 & 0.76 & 0.97 & 0.99 & 0.45 & 0.36 & 0.94 & 0.66\\
FLAME-100 & 0.84 & 0.66 & 0.24 & 0.35 & 0.27 & 0.44 & 0.98 & 0.74 & 0.97 & 0.99 & 0.38 & 0.31 & 0.93 & 0.65 \\ 

\bottomrule
\end{tabular}

\begin{tablenotes}
\item[1] Number preceding `-' presents respective FLAME$_{Perturb}$ test case. For instance, FLAME-25 presents results for methods trained on FLAME and tested on FLAME$_{25}$
\item[2] WriteprintsRFC; \item[3] RoBERTa-multi ; \item[4] DeBERTa-Finetuned; \item[5] T5-sentinel;
\end{tablenotes}

\end{threeparttable}
\label{tab:Attack_AA}%
\end{table*}
}

\begin{document}
% tracking the digital forensics and incident response from LLM perspective.
% how prepared is digital forensics and incident response for LLM-based attacks
% \title{Text-Based DFIR: Uncovering the Gaps in Neural Text Detection and Forensic Authorship Attribution}
\title{Is the Digital Forensics and Incident Response Pipeline Ready for Text-Based Threats in LLM Era?}
% Towards Robust Text-Based Digital Forensics: 
% \title{Revisiting the DFIR Pipeline: Addressing Detection and Authorship Attribution Issues in Text-Based Security Systems}
% Text-Based DFIR: Uncovering the Gaps in Detection and Authorship Attribution
% Evaluating the DFIR Pipeline for Text-Based Security: Challenges in Detecting Machine-Generated Text and Attributing Authorship

\author{Avanti Bhandarkar, Ronald Wilson, Anushka Swarup, Mengdi Zhu, Damon Woodard

Florida Institute for National Security, 

University of Florida

% % <-this % stops a space
\thanks{Avanti Bhandarkar, Ronald Wilson, Anushka Swarup, Mengdi Zhu, and Damon Woodard are with the Florida Institute for National Security, University of Florida, Gainesville, Florida, USA-32611 (email: \{avantibhandarkar, ronaldwilson, aswarup, zhum, dwoodard\}@ufl.edu)}}
% <-this % stops a space



\maketitle

\begin{abstract}

In the era of generative AI, the widespread adoption of Neural Text Generators (NTGs) presents new cybersecurity challenges, particularly within the realms of Digital Forensics and Incident Response (DFIR). These challenges primarily involve the detection and attribution of sources behind advanced attacks like spearphishing and disinformation campaigns. As NTGs evolve, the task of distinguishing between human and NTG-authored texts becomes critically complex. This paper rigorously evaluates the DFIR pipeline tailored for text-based security systems, specifically focusing on the challenges of detecting and attributing authorship of NTG-authored texts. By introducing a novel human-NTG co-authorship text attack, termed CS-ACT, our study uncovers significant vulnerabilities in traditional DFIR methodologies, highlighting discrepancies between ideal scenarios and real-world conditions. Utilizing 14 diverse datasets and 43 unique NTGs, up to the latest GPT-4, our research identifies substantial vulnerabilities in the forensic profiling phase, particularly in attributing authorship to NTGs. Our comprehensive evaluation points to factors such as model sophistication and the lack of distinctive style within NTGs as significant contributors for these vulnerabilities. Our findings underscore the necessity for more sophisticated and adaptable strategies, such as incorporating adversarial learning, stylizing NTGs, and implementing hierarchical attribution through the mapping of NTG lineages to enhance source attribution. This sets the stage for future research and the development of more resilient text-based security systems.
\end{abstract}


\section{Introduction}

Recent developments in Artificial Intelligence (AI), especially in Generative AI (GenAI) through large language models (LLMs), have significantly improved natural language generation (NLG) capabilities. These models, also known as Neural Text Generators (NTGs), like OpenAI's ChatGPT, excel in tasks ranging from creative writing to coding, producing high-quality text indistinguishable from human writing \cite{gehrmann2019gltr,ippolito2020automatic,wu2023survey}. Despite these benefits, GenAI also poses risks, notably by creating realistic deepfakes and facilitating cybercrimes such as phishing, disinformation campaigns, and denial of service (DoS) via business email compromise \cite{gupta2023chatgpt}. These threats range from low-risk activities like spamming to high-risk targeted attacks that threaten personal, corporate, and national security. Spear phishing emails, in particular, remain the primary and prevalent attack vector utilized by many Advanced Persistent Threats (APTs) to compromise victims' computers \cite{urban2020plenty}.
A recent IBM X-Force study highlights the ease of crafting sophisticated phishing emails with ChatGPT, raising concerns among law enforcement about the potential rise in cybercriminal activities due to advanced LLM technology \cite{ibmX,europol2023,fbi2023}.

Given the potential for attacks using NTGs, considerable research has focused on Neural Text Detection (NTD) \cite{uchendu2023attribution,jawahar2020automatic,crothers2023machine}. 
This line of research often operates under two assumptions: firstly, that detecting NTG-authored text alone can safeguard privacy and security, and secondly, that attackers employ solely NTG-authored unperturbed text.
While these assumptions may hold for low-risk scenarios, in high-risk situations the attackers may possess advanced knowledge, motivation, and techniques to evade detection jeopardizing the privacy and security of the defense systems. In such cases, in addition to NTD, a comprehensive forensic analysis of the text is imperative to identify the identities of NTG and human attackers to pinpoint attack vectors and vulnerabilities in defense systems. 

Furthermore, research suggests that minor perturbations resulting from word and sentence substitutions can cause misclassifications by NTDs \cite{shi2024red,zhou2024humanizing}.
However, a less explored but naturally occurring perturbation arises from human-NTG co-authorship, where attackers modify NTG-authored text for malicious purposes. Open-source LLMs, which require seed text for generation, are particularly susceptible to such manipulations. Although commercial NTGs like ChatGPT have safeguards for ethical use, attackers can still exploit these models to generate benign text initially and then inject malicious content, creating a blend of human-NTG co-authored text.

Although the use of AI in digital forensics is generally limited, the classical cybersecurity framework, Digital Forensics and Incident Response (DFIR) provides a systematic framework for evaluating the defense systems \cite{dunsin2024comprehensive}. Adapted for text-based attacks, as illustrated in Figure~\ref{fig:dfir}, the DFIR pipeline enables attack detection and prevention while concurrently identifying the identities of both humans and NTGs.
As previously noted, most studies in this field focus solely on the incident detection phase where NTD serves as the core component, while neglecting the forensic profiling phase. Although Authorship Attribution (AA) has been effectively employed to attribute texts to human authors, the forensic attribution of texts authored by NTGs, and the subsequent forensic profiling of texts in text-based attack scenarios, are areas that remain largely unexplored.

This gap in research restricts the depth of insights into adversary tactics, techniques, and procedures (TTPs) that can be derived from text data, thus limiting its potential. 
Furthermore, in real-world scenarios with high risks and high rewards for attackers, the text is expected to be poisoned or perturbed through the human-NTG collaboration. These aspects, crucial for understanding and mitigating risks, have been minimally explored and even less often studied in conjunction.

Contrary to the prevailing research focus on distinguishing between human and NTG text in ideal scenarios where the text is solely authored by either human or NTG, this study raises two interconnected questions: ``\emph{Does the ideal scenario accurately mirror real-world conditions?}" and ``\emph{Can we reliably detect and attribute the sources of human and NTG identity in the presence of adversarially co-authored text?}"

The rest of the paper is organized as follows: Section 2 discusses the related works that support the evaluation of the DFIR pipeline. Section 3 details the human-NTG co-authorship attack -CS-ACT- and provides heuristics for evaluation under two scenarios - ideal and adversarial attack. Section 4 discusses the results in detail while Section 5 aims to understand the challenges for the weakest link in the DFIR pipeline- NTG-AA. Finally, Section 6 provides directions for future research motivated by our comprehensive evaluation and the work is concluded in Section 7.

\noindent\textbf{Key Contributions:}
\begin{itemize}
    \item First comprehensive examination of the DFIR pipeline for text-based attacks, addressing NTD, NTG-AA, and human-AA using 22 algorithms, across 14 datasets and 43 unique NTGs in both controlled and real-world scenarios.
    \item A novel adversarial attack, \emph{CS-ACT}, involving human-NTG co-authorship is introduced. Alongside this, the \emph{FLAME} dataset is presented, featuring 25 unique NTGs including the latest GPT-4o, with varying levels of co-authorship. This dataset will be publicly available to facilitate further research on adversarial attacks.
\end{itemize}

\DFIR

\section{Related Works}

In the context of cyberattacks involving textual data, we delineate the DFIR pipeline, depicted in Figure~\ref{fig:dfir}, into three distinct phases: Text Generation, Incident Detection (proxy for IR), and Forensic Profiling (proxy for DF). 

\subsection{Text-Based Attacks}

Text-based attacks on security systems involving NTGs aim to manipulate the text generation process, with successful infiltration defined by the ability of the altered text to induce misclassifications by either the NTD, AA, or both.
These attacks can be broadly categorized into two types: train-time and inference-time attacks. Train-time attacks involve the manipulation of training data or learning infrastructure to undermine the model's security and efficacy such as backdoor attacks \cite{chen2021badnl,pan2022hidden}. However, these attacks are resource-intensive, time-consuming, and require control over the training process of the defense mechanism - a requirement attackers may struggle to consistently fulfill, particularly in secure settings.

On the other hand, a more feasible scenario is presented by inference-time attacks, wherein the attackers modulate the output of the NTG to produce text that can cause misclassifications. 
Most recent studies have reported the effectiveness of attacks using perturbed text to induce misclassifications where the majority of research is focused on attacking NTDs, with little to no attention given to the more challenging task of source attribution from the perturbed text.
Recent studies on text-based attacks typically involve surface-level perturbations like word or sentence substitutions \cite{shi2024red,zhou2024humanizing}. Others opt for transforming entire texts through paraphrasing \cite{sadasivan2023can,krishna2024paraphrasing}.
However, there has been little to no exploration into how varying degrees of collaboration between humans and NTGs affect both NTD and AA tasks.

\subsection{DFIR Infrastructure}

NTD and AA are fundamental components of the DFIR infrastructure. NTD methods differentiate between texts generated by humans and NTGs, framing this as a binary classification problem. Extending this further, AA aims to identify the specific author of the text from a pool of potential authors, posing a more complex multi-class classification challenge.

While NTD has garnered extensive research attention due to its lower complexity, leading to the development of numerous sophisticated algorithms, AA research has primarily focused on identifying human authors, with limited progress in identifying NTGs. However, the methodologies employed in both NTD and AA can generally be categorized into three main types based on the features they extract and the classifier architecture they utilize: \emph{metric}-based, \emph{content}-based, and \emph{model}-based approaches.

\detectorTable

\emph{Metric}-based methods center on language compositionality and compute predefined statistics or metrics from text data, such as entropy, perplexity, or log-likelihood.
Typically, formulated as a statistical outlier detection problem, these methods employ a zero-shot approach using simple binary threshold or train a basic machine learning classifier on the statistical features \cite{mitchell2023detectgpt,frohling2021feature,gehrmann2019gltr}. Generally employed for NTD, these methods are often promoted as hassle-free as they do not necessitate training on complex ML and DL architectures. We employ seven most popular metric-based methods for NTD.

\emph{Content}-based methods utilize NLP techniques to extract linguistic and content-related features from the text, which are then applied to downstream ML and DL algorithms such as Decision Trees and Neural Networks \cite{aich2022demystifying,abbasi2008writeprints}. Prominent features such as the Writeprints feature set, psycholinguistic features like LIWC, and linguistic metrics like readability have gained traction in AA and have been adapted for NTG-AA tasks \cite{abbasi2008writeprints,uchendu2020authorship, aich2022demystifying}.
N-grams, varying across different language levels such as character, word, and parts-of-speech (POS), are another cornerstone in AA and Stylometry \cite{kevselj2003n, stamatatos2012robustness, kestemont2018overview}. These features are commonly paired with machine learning classifiers like Support Vector Machines (SVM) to determine authorship \cite{kestemont2019overview, sari2017continuous}. Notably, character n-grams, especially 2 and 3-grams, have proven effective in capturing the subtle nuances of an author’s style, occasionally outperforming advanced algorithms \cite{wilson2021sqse}.
Generally, the content-based methods aim to encapsulate aspects of a text’s organization, logical structure, formality, style, and objectivity \cite{munoz2023contrasting,guo2023close}. We employ four content-based methods, where three, LIWC, WRFC, and LFI are applied to all three tasks while character-3-grams is applied to the AA tasks.

\emph{Model}-based methods represent the state-of-the-art (SOTA) across three key tasks, utilizing advanced deep learning architectures often in conjunction with fine-tuning transformers like BERT, RoBERTa,  DeBERTa, etc. or other prominent LLMs such as XLNet \cite{solaiman2019release,uchendu2021turingbench,munir2021through,rosati2022synscipass}. 
Integrating stylistic features with fine-tuned pre-trained language models like BERT and DeBERTa has shown promising results in human AA \cite{fabien2020bertaa, barlas2020cross}. 
Additionally, researchers have experimented with modifying the learning architecture by either incorporating a specialized neural network like CNN on top of the fine-tuned LLM or employing different learning objectives such as contrastive learning for human and NTG identification \cite{ai2022whodunit,diwan2021fingerprinting}.
This approach effectively merges elements from both metric and content-based methods. Transformer models capture sequence-based language composition information akin to metric-based methods due to their bidirectionality and enrich this with contextual insights and self-attention mechanisms, critical for NTD tasks. Moreover, the integration of stylistic features adds discriminability, enhancing AA capabilities. We utilize nine representative SOTA model-based methods. AIGC is dedicated exclusively to NTD tasks. Four methods—ContraX, BertAA, DeBERTa-FT, and XLNet-FT—are specifically employed for AA tasks. Meanwhile, T5-Sentinel, BERT-FT (BERT-multi), and openAIDetector (RoBERTa-multi) are applied to both NTD and AA tasks.

A brief description of these methods is provided in Table~\ref{tab:detectors}

\section{Compromising Security via Adversarially Co-authored Text (\textit{CS-ACT})}

We present the CS-ACT attack task designed to undermine the security and privacy of systems by targeting the DFIR pipeline. The attacker(s) aim to disrupt this pipeline by producing adversarially co-authored text, crafted with inputs from both human and NTG identities. This introduces errors in both the incident detection and forensic profiling phases, compromising the system's overall security measures.

\noindent \textit{Problem Statement}:

In a security system employing the DFIR pipeline, the attacker has two objectives. The first goal is to discreetly circumvent the Incident Detection phase without setting off any alerts. Since this phase is designed to detect NTG-authored text, the attacker successfully meets this goal if they can sufficiently confuse the NTD system into believing that the adversarially co-authored text is entirely human-authored, thereby evading detection.

Should the attack — involving adversarially co-authored text — be accurately detected in the initial phase, the attacker's secondary objective shifts to disguising their writing style. This tactic is designed to introduce ambiguity during the Forensic Profiling phase, thereby hindering the precise identification of the text’s true authorship for both NTG and human identities.

\subsection{Adversarially Co-authored Dataset Creation}

Given the new challenge of exploring the impact of human-NTG co-authorship on the DFIR pipeline, we curate a new dataset, named, \emph{Forensic LLM-Authorship Mixture Evaluation (FLAME)} featuring 25 distinct NTGs including autoregressive and chat models. This dataset is divided into two subsets: \emph{FLAME$_{Pure}$}, which includes texts authored solely by humans or NTGs, and \emph{FLAME$_{Perturb}$}, which contains texts that are adversarially co-authored. FLAME$_{Pure}$ follows the text generation protocol found in existing research and serves as a basis for comparison with other publicly available datasets while FLAME$_{Perturb}$ is developed to include adversarially co-authored text, aligning with the specific requirements of the CS-ACT attack scenario.

The initial pool of authors was selected from the Blogs Authorship Corpus (BAC). The pre-processing steps included removing duplicate spaces, line breaks, non-ASCII characters, initial punctuations, and specific tokens like ``urlLink''.
Only authors who had at least 10 text samples exceeding 200 tokens and could provide at least 125 non-overlapping chunks of 400 tokens after pre-processing were retained, narrowing down to 135 authors. A flowchart of the dataset creation process and organization for FLAME is depicted in Figure~\ref{fig:flame}. This process is repeated for every NTG-Human author pair. Additional details including a list of NTGs, implementation details, etc. are in Supplementary materials.

For human-AA, we included authors who contributed more than 100 samples each to FLAME$_{Human}$, resulting in a total of 97 authors. To manage class imbalance for this task, the number of samples per author was capped at 125.

Unlike FLAME$_{Pure}$, where the seed human text (H$_{seed}$) is removed after text generation, all subsets of FLAME$_{Perturb}$ retain both the H$_{seed}$ and the NTG-generated text (N$_{gen}$) except for FLAME$_{0}$ which mirrors the pure subset with the first 50 tokens used as H$_{seed}$. Conversely, FLAME$_{100}$ represents purely human-authored text in the form of original author text (H$_{seed}$=N$_{gen}$). All text data generated from the NTGs was pre-processed post removal of H$_{seed}$.

\FLAMEOrg

This dataset makes the first and the sole dataset offering concurrent human and NTG author labels along with NTG-human co-authored samples. 

\subsection{Evaluation Scenarios} \label{sec:eval_scenarios}

In many NLP applications, it is often observed that even highly proficient models exhibit sensitivity towards specific data properties such as topic, domain, genre, etc. \cite{al2017approaches,barlas2020cross,bhandarkar2023navigating,waldis2024dive}. Consequently, these models tend to perform better when the data shares similar properties. However, in real-world attack scenarios, controlling these data properties is seldom feasible. Therefore, relying solely on evaluations conducted in controlled setups where data properties remain consistent may not suffice to ensure a reliable assessment of NTD and AA methods. 
Therefore, in this study, we assess the SOTA methods on three levels - the first two, \emph{within-dataset (WD)} and \emph{across-datasets (AD)}, operate under ideal conditions, while the third addresses \emph{CS-ACT attack} scenario - each representing a progressively higher level of complexity.

To emulate the ideal scenario - text solely authored by humans or NTGs - we incorporate six widely used publicly available datasets. These datasets collectively exhibit diverse textual characteristics, including varying numbers of NTGs, text generation techniques, source platforms, sample counts, and sample lengths. For NTD, samples for all NTGs collectively represent the ``NTG" class whereas for NTG-AA, each NTG is treated as a distinct author.
The summary statistics of all datasets used in this study are presented in Table~\ref{tab:data_stats}.

\dataStats

\emph{WD} evaluations are carried out independently for each dataset to assess the vulnerabilities of SOTA methods to specific data characteristics and NTGs. However, since these datasets are primarily designed for a Turing-test style NTD, some of them exhibit a natural bias towards a higher number of NTG samples, sometimes leading to a class imbalance. 

\emph{AD} assessment evaluates the vulnerabilities of NTD and AA methods to variations in data characteristics and number of NTGs. To accomplish this, we create a comprehensive super-set by merging all datasets and forming independent super-sets for NTD and NTG-AA tasks. In the absence of human author identities in the publicly available datasets, we exclusively use the FLAME dataset for human-AA. Therefore, the AD evaluation does not apply to human-AA task.

For NTD, we collate text samples from each dataset to represent each NTG variant equally, accumulating 1000 samples for every variant. For example, the model bloom-7b1, featured across 11 datasets, contributes approximately 91 text samples per dataset. A corresponding number of human-authored samples are selected from every dataset to create a balanced NTD super-set.

Next for NTG-AA, we combine the NTGs with varying parameters from the NTD super-set to establish a base model representation. For instance, the Llama2 base model includes all parameter variants, such as Llama2-7B, Llama2-14B, Llama2-65B, etc. This approach is based on findings from multiple studies suggesting that NTG variants of the same base model tend to share similarities in text generation \cite{pagnoni2022threat,antoun2023text}. An equal number of samples from each dataset are then pooled, ensuring 1000 samples for each NTG base model.

For both \emph{WD} and \emph{AD} scenarios, the NTD and AA methods are trained and evaluated using text that is either entirely human or NTG-authored. 
However, the \emph{CS-ACT} attack scenario exclusively tests the methods against adversarially co-authored texts using the FLAME$_{Perturb}$ while training is conducted on WD (FLAME$_{Pure}$) and AD scenarios respectively. To ensure generalizability and consistency, five-fold cross-validation is employed for all experiments.

\subsection{Performance Metrics}

In an ideal scenario, it is crucial to evaluate model performance across all target classes to ensure comprehensive effectiveness. This evaluation commonly employs the Macro-F1 score, which computes the F1 score for each class individually and then averages these scores, providing equal importance to every class. This method ensures that no single class disproportionately influences the overall model performance. Given the sequential nature of the DFIR pipeline, confirming robust performance across all classes in controlled environments is essential. Therefore, we average the macro-F1 scores across all cross-validation folds to assess the robustness of the models.

Conversely, in adversarial attack scenarios, the Attack Success Rate (ASR) is the preferred metric. ASR quantifies the proportion of successful attacks, providing a direct measure of the model's vulnerability \cite{yao2024survey,shi2024red}. Consistent with this approach, we utilize ASR to assess the robustness of NTD and AA methods under the CS-ACT attack scenario. 

For AA, we define ASR as the average proportion of misclassifications per class.
For NTD, within the context of CS-ACT attack, every test case is regarded as a potential attack, irrespective of the human contribution to the NTG-human co-authorship, except FLAME$_{100}$. Consequently, we adjust the ASR to reflect the proportion of test samples that are misclassified as human-authored. ASR calculations for both NTD and AA are given in Equation~\ref{eq:asr_ntd}. The final ASR for each experiment is the average ASR across all cross-validation folds.

\begin{equation} \label{eq:asr_ntd}
        \text{ASR}_{NTD} = \frac{m_h}{T}; \text{ASR}_{AA} = \frac{1}{N} \sum_{i=1}^{N} \frac{m_i}{t_i}
\end{equation}

where,
\begin{itemize}
\item[$m_h$] is the number of NTG-authored samples predicted as human-authored.
    \item[$T$] is the total number of test samples.
    \item[$N$] is the total number of classes (representing the number of NTGs or human identities),
    \item[$m_i$] is the number of misclassified instances in class $i$ (where the predicted label does not match the true label),
    \item[$t_i$] is the total number of instances in class $i$.
\end{itemize}

\section{Results}

Given the sequential structure of the DFIR pipeline, each phase significantly influences the subsequent phase. This becomes especially pronounced when handling adversarially co-authored text, rendering the entire pipeline vulnerable following the initial phase. The perturbed text may cause misclassifications during the NTD phase that can compromise system security and privacy, making subsequent phases unreliable. Even if the perturbed text is accurately detected by an NTD, it can still result in misclassifications by the AA methods, enabling attackers to effectively evade identification. Therefore, it is essential to independently explore the vulnerabilities associated with each of these phases.

\detectASR

\subsection{Ideal Scenarios}

Results from the ideal scenario are illustrated in Figure~\ref{fig:ideal_results} using a box-and-whisker plot, with the spread of the box reflecting the impact of dataset-specific characteristics on a specific algorithm. Comprehensive results for each dataset and algorithm are provided in Tables ~\ref{tab:ASR_detect} and ~\ref{tab:NTGAA_ASR} for the incident detection and forensic profiling phases, respectively.

\noindent\textbf{NTD: }

In NTD, three key insights have been derived. First, in line with previous studies, the results confirm the robustness of model-based methods for NTD tasks in WD and AD scenarios. However, the effectiveness of these methods diminishes in scenarios featuring diverse data characteristics, such as those found in the AuTexTification dataset and AD evaluations. In comparison to AuTexTification, a higher performance of these methods on AD scenario suggests that the presence of a greater number of NTGs and varied data characteristics may enhance algorithmic generalization.

Secondly, the content-based method that utilizes simple stylistic features, such as WRFC, perform on par with its more sophisticated and resource-intensive model-based counterparts, and it maintains stability for both WD and AD scenarios. This indicates that significant stylistic differences likely exist between human and NTG-authored texts.

Finally, contrary to findings from other research, metric-based methods show the weakest performance with the highest variability, indicating their sensitivity to specific dataset characteristics. Moreover, methods such as detectGPT and detectLLM, underperform on datasets with older NTGs like TweepFake, Uchendu, and TuringBench. This lack of effectiveness and generalizability in broader real-world applications indicates the unreliability of these methods for NTD.

In the AD scenario, which encompasses samples from 73 NTGs, including unique models and variants of base models, a maximum macro-F1 score of 0.93 highlights notable security levels but also points to significant areas for improvement. These observations suggest that even SOTA NTD systems can be susceptible to specific data characteristics and NTG properties, potentially compromising the overall security framework.

\IdealResults

\noindent\textbf{Human-AA: }
Two key insights from the human-AA results emerge - the significance of subtle nuances and the limitations of methods solely reliant on contextual understanding of content in accurately discerning between human authors.

Consistent with the literature in traditional AA, the character n-gram method achieves the highest performance for human-AA, surpassing more advanced model-based features. 
This can be attributed to the evolution of human writing style through experiential processes, resulting in subtle distinctions between authors as a result of individual habits and life experiences. Characters, being the smallest textual units, aptly capture these nuanced stylistic differences. On the other hand, the lowest performance is observed for model-based approaches utilizing RoBERTa and similar methods, which prioritize contextual comprehension of content. 
Despite excelling in the NTD task, the reduced performance of these methods for human-AA suggests that solely relying on generated content is inadequate for distinguishing human authors.

Conversely, model-based approaches that integrate language compositionality with content analysis demonstrate effectiveness, as evidenced by the notably higher performance of methods such as ContraX and Deberta-V3. Moreover, given that XLNet is an autoregressive NTG, the high performance by XLNetFT suggests that human authors also demonstrate diversity in the sequential compositionality of language.

Compared to NTD, human-AA presents a more intricate challenge, prompting concerns regarding the accurate and reliable attribution of attacker identities during the forensic profiling phase.

\NTGAA

\noindent\textbf{NTG-AA: }
In the WD scenario, significant performance variation is observed across various datasets, underscoring the challenge of recognizing specific NTGs with distinct data characteristics. 

Comparing the performance of algorithms in WD versus AD scenarios reveals that while the average performance is comparable, generalizability at the NTG level decreases when exposed to diverse data characteristics. To verify this, a simple statistical test was employed on the AD evaluations. The Spearman rank correlation was calculated between the recall values for each NTG and the number of NTGs that contain samples from more than two datasets. A statistically significant negative correlation of -0.57 was obtained suggesting that an NTG's performance generally declines in the presence of varied data characteristics. 

Secondly, datasets containing NTG variants consistently show low performance across all algorithms. A comparison between all DFTD subsets and the FLAME dataset, which contain roughly the same number of NTGs, further illustrates this point; performance on FLAME is significantly higher than on DFTD. Moreover, the average performance for individual datasets is comparable to, and at times lower than, performance in scenarios that involve combined base models across datasets. These findings lend further credence to the notion that text generation by NTG variants exhibits similar characteristics.

\noindent\textbf{Comparing NTG-AA and Human-AA: }
An inverse relationship is observed between the best-performing algorithms for NTG-AA and Human-AA, highlighting distinct performance trends for each problem. This indicates the distinct nature of these two problems necessitating tailored solutions, as the challenges of human-AA cannot be addressed in the same manner as that of NTG-AA.

While character n-grams, commonly associated with stereotypical style features, exhibit strong performance for human-AA, they perform less efficiently for NTG-AA in comparison. Conversely, model-based features leveraging contextual understanding of content, which perform suboptimally for human-AA, excel in NTG-AA. These findings reinforce the notion that the concept of ``style'' differs between humans and NTGs. Based on these results, it appears that distinguishing between NTGs relies on ``what'' content they generate, whereas the distinctions among humans stem from ``how'' the content is generated. Specifically, in the context of NTG-AA, the ``what'' may correspond to choices in NTG architecture, the data used for training, knowledge injection strategies, and other factors, whereas the ``how'' may be influenced by decisions made during text generation, such as decoding strategies.

However, we observe that certain algorithms, which consider a blend of both content and structure and/or stylistic aspects of text generation, demonstrate comparable performance across both tasks. This is evident in the top-performing method in NTG-AA and the second-best in human-AA — ContraX.
This suggests that methods capable of addressing AA for both NTGs and humans concurrently must integrate elements such as content, context, linguistic structure, and style to effectively perform these tasks under ideal scenarios.

Assuming that each NTG and human identity produces language with distinctive styles, NTG-AA (for 43 NTGs) significantly underperforms compared to human-AA (for 135 authors). 
This indicates two possibilities: either the NTGs do not possess unique stylistic attributes or the current SOTA algorithms cannot differentiate between them.
Nevertheless, in comparison to both NTD and human-AA, NTG-AA represents the most challenging problem in the entire DFIR pipeline.

\AttackResults

\subsection{CS-ACT Attack Scenarios}

\detectAttack

Attacks are typically evaluated on ``secure'' systems that demonstrate a certain level of efficiency under ideal scenarios. In the absence of established heuristics for efficiency thresholds, we consider methods with a macro F1 score greater than 0.5 in the ideal scenario as ``valid'' (or secure) for evaluating the CS-ACT attack. Therefore, the results for different attack scenarios, which involve varying proportions of human-NTG co-authorship as discussed in Section~\ref{sec:eval_scenarios}, are presented in Figure~\ref{fig:attack_results}. Detailed results for all attack scenarios are provided in Tables~\ref{tab:detect_attack} and ~\ref{tab:Attack_AA}. To facilitate comparison across the various problems addressed in this study, the results are illustrated using box-and-whisker plots for each perturbation scenario of $P=p$ where the spread of the box represents variation in performance across algorithms.

\noindent\textbf{Effect on Incident Detection: }
We observe that NTD algorithms are susceptible to the CS-ACT attack. Even minimal perturbation ($P=25$) nearly doubles the average ASR. As the proportion of human-authored content increases, the ASR continues rising exponentially, particularly in the WD scenario.

This susceptibility is especially pronounced in model-based methods, likely due to their tendency to overfit purely NTG-generated text. As a result, these methods may interpret the addition of human content as noise and ignore it. This issue is most evident at the highest level of human-author co-authorship ($P=75$) in the WD scenario, where these methods yield an average ASR of 0.83. This also implies that there must be greater similarity in the NTG-authored text compared to human-authored text, helping explain why human-AA typically outperforms NTG-AA in the ideal scenarios. 

On the contrary, content-based features appear to yield a lower valid ASR, likely due to their shared ability to capture stylistic or psycholinguistic aspects of text, including sentiment and emotion. For instance, the next best performing method in ideal scenarios, WRFC, exhibits better performance than model-based methods but similarly shows an exponential rise in ASR with increased human authorship. This can be attributed to the challenge of style disentanglement from traditional AA. When a text is co-authored, the representative features become a composite of both authors, entangling their styles and leading to misclassifications.

\AAAttack

Comparing the two scenarios, it is clear that the AD scenario exhibits a lower ASR, suggesting that training systems on a broader range of NTGs and diverse data characteristics enhance security, albeit not perfectly. This supports the strategy of incorporating a more extensive array of NTGs and varied data attributes to train NTD systems more effectively.

Yet, these findings fall short of meeting the ideal security standards, underscoring the vulnerability of the defense system to CS-ACT attacks. In addressing the earlier posed research question, \emph{it appears that the reliability of the incident detection phase remains insufficient}.

\noindent\textbf{Effect on Forensic Profiling: }
In this phase, NTG-AA is particularly impacted by CS-ACT attacks where under the minimal perturbations, more than half of the samples are misclassified producing an ASR of 0.60 and 0.76 respectively for WD and AD scenarios. While the highest valid ASR for human-AA under maximum perturbation (P=25) is capped at 0.73, for NTG-AA it reaches 0.96 for WD evaluations. Human-AA also displays greater variability in ASR under high perturbations, indicating that some methods remain effective for this task even under severe conditions - a resilience not as evident for NTG-AA. Despite this, the notably high ASRs under minimal perturbations—0.3 for human-AA (at P=75) and 0.6 for NTG-AA (at P=25)—underscore the overall susceptibility of the forensic profiling phase to CS-ACT attacks.

While human-AA shows consistency with the ideal scenario, where top-performing methods like character n-grams and ContraX still perform well under attack, the same cannot be said for the top-performing methods for NTG-AA. This reinforces the critical distinction in style between humans and NTGs mentioned earlier. For instance, consider ContraX, which performed well for both human and NTG-AA in the ideal scenario. The representations obtained from fine-tuning the transformer with a contrastive objective results in more resilient style capture for humans. However, this method shows one of the highest valid ASRs for NTG-AA under high perturbation. This is because, in the absence of clear discriminability in styles for NTGs, the base representations for each NTG learned by the algorithm are not robust.

As previously mentioned, the concept of style for NTGs appears to be embedded in the content and compositionality of the text. This is supported by the superior performance of T5-Sentinel for NTG-AA under attack scenarios. 
By sequentially predicting tokens, T5-Sentinel more effectively discerns the subtle differences in content and structure used by different NTGs, thereby enhancing its performance. While these results suggest potential heuristics for developing more robust NTG-AA algorithms, the ASR greater than 0.6 under minimal perturbations in both WD and AD scenarios makes it challenging to draw definitive conclusions from these results.

% Rate of increase
From Figure~\ref{fig:attack_results}, it is evident that the ASR rapidly reaches a saturation point, approaching the maximum achievable ASR for NTG-AA, while the minimum ASR in all scenarios surpasses that of human-AA by a significant margin. Considering that even under ideal conditions, NTG-AA poses a greater challenge makes it the weakest link in the DFIR pipeline, followed by NTD, and then human-AA. In response to the research question, it is evident that \emph{the overall reliability of the forensic profiling phase is significantly undermined by the CS-ACT attack scenario, particularly concerning NTG-AA}.

\section{Challenges in NTG-AA}

With the rapid proliferation of NTGs, it is imperative to assess why NTG-AA emerges as the most challenging issue in the DFIR pipeline, thereby hindering reliable forensic profiling. To understand this, we analyze NTG-AA performance in the AD scenario at the NTG level (detailed results in Supplementary material). This is illustrated using a box-and-whisker plot, where the spread of the box represents the variability in performance due to different algorithms, as depicted in Figure~\ref{fig:LLMwiseNTGAA_across}. Two key insights can be gained relating to \emph{model sophistication} and \emph{distinctiveness of style in NTGs}. 

First, NTGs with the highest attribution performance tend to be older models, while newer NTGs perform the worst. This suggests that model sophistication is a significant factor in the challenges of attributing NTGs, albeit with exceptions such as GPT4, GPTNeoX, Phi3, etc. Second, treating NTGs as distinct identities despite a common base architecture, such as variants of the proprietary OpenAI GPT-3 models (e.g., Babbage, Curie, text-davinci), might lead to lower attribution performance when assessed individually. Furthermore, despite sharing a common base architecture (GPT-3), open-source models like BTLM, GPT-Neo, and MPT exhibit significantly lower attribution performance. This discrepancy could be attributed to model size, as proprietary models are typically larger while open-source models are smaller, indicating that model size might also influence NTG style.
Therefore, it is plausible that the distinctiveness or commonalities of style resulting from the base architecture play a crucial role in the identification of NTGs.
Therefore, two lines of inquiry are needed: exploring the impact of \emph{model sophistication} and \emph{distinctiveness of NTG styles} on NTG-AA performance.

\LLMwiseNTGAA

\subsection{Model Sophistication}

Model sophistication can be examined from various perspectives. First, we use the year of a model's release as a proxy for its sophistication. Second, we consider the quality of text produced by an NTG as an indicator of sophistication. Lastly, we assess a model's ability to produce human-like text as a measure of its sophistication. 

\subsubsection{Year of Release}

We calculate the Spearman rank correlation ($R$) between the year of release and the average performance of all NTGs from the pre-LLM era (before 2018) including RNNs and Markov Models, to 2024 including models like Gemma and GPT-4o. We find a strong negative correlation of -0.83, with a statistical significance of less than 0.01. This suggests that NTG-AA performance decreases with an increase in model sophistication.

Since 2018, NTG-AA performance consistently declined until 2024, when the GPT-4o model led to notable improvement. This improvement is likely a response to potential misuse of genAI technology, prompting NTG developers to embed subtle ``watermarks'' in the text that are undetectable by humans but recognizable by algorithms \cite{kirchenbauer2023watermark}. It is plausible that recent models like GPT-4, GPTNeoX, Phi3, and GPT3.5 utilize this technology, enhancing security and contributing to their comparatively better identification. However, \emph{the consistent decline in NTG attribution with model sophistication continues to be a significant concern}.

\subsubsection{Text Quality}

We utilize three quality metrics to evaluate text quality—coherence, repetition, and grammaticality—similar to the GRUEN metric \cite{zhu2020gruen}. Additionally, we incorporate an assessment of stylistic quality using the SQSE metric \cite{wilson2021sqse}. The metrics are described below:

\begin{enumerate}
    \item Grammaticality score: This metric assesses sentence likelihood and grammar acceptance by employing both raw and fine-tuned BERT on the Corpus of Linguistic Acceptability (CoLA). We utilize the implementation from GRUEN for computing this score. A higher score indicates better grammaticality.
    \item Coherence score: This metric calculates the Sentence-Order Prediction (SOP) Loss between the correct and swapped elements in the text. Following GRUEN, two variants at the sentence and word levels were derived. A higher score indicates better word or sentence coherence.
    \item Redundancy score: This metric assesses the likelihood of duplicate content. It analyzes the frequency of word n-grams ($1\geq n\geq5$) and calculates the proportion of unique n-grams occurring multiple ($k$) times ($k \in [2,6]$). The final score is calculated as the negative mean of the logarithms of these proportions, aiming to penalize repetitions in higher-order n-grams. A low redundancy score indicates text filled with unique words, while a high score indicates exact word or phrase repetitions.
    \item SQSE score: This metric assesses the stylometric quality of a text sample by extracting part-of-speech (POS) bigrams, character unigrams, and word length distribution. A higher score indicates a greater likelihood of the presence of a distinctive writing style.
\end{enumerate}

First, we extract quality metrics from NTG-AA super-set text samples. Next, we calculate Spearman's rank correlation ($R$) between the average macro-F1 for each NTG and the average metric for that NTG. The results are presented as Spearman-$R$ in Table~\ref{tab:text_quality}. Since SQSE is designed to evaluate the stylistic quality of text for human authors, we observe an insignificant correlation, indicating that there is little to no relationship between stylistic quality and NTG-AA performance.

\EffectTextQuality

For linguistic quality metrics, we observe a negative correlation with NTG-AA performance. The weaker significance and lower correlation for redundancy can be attributed to the metric's nature, where extremes signify poor text quality. Even humans tend to repeat words and phrases, so some redundancy is expected in good-quality text. Earlier observations showed a negative correlation between NTG-AA performance and model sophistication. Linking these results to text quality suggests that recent NTGs are capable of better language production. Overall, \emph{higher text quality provides anonymity to NTGs, raising concerns for reliable attribution}.

\subsubsection{Likeness to Humans}

Research suggests that factors such as fluency, repetitions, typos, grammatical errors, and coherence contribute to making machine-generated content appear human-like \cite{jawahar2020automatic}. To evaluate the similarity of NTG-authored content to human writing, we use the ``human'' class samples from the NTD super-set and calculate the mean and standard deviation of the distribution for each metric discussed above. Subsequently, for each NTG, we compute the Z-scores using the computed mean and standard deviation for each specific metric.
The z-scores assess the dissimilarity between the metrics from texts authored by NTGs and the overall human distribution. Lower z-scores indicate that the NTG produces text closer to what humans do.
Finally, Spearman-$R$ is calculated between the NTG-AA performance and the average z-score for each NTG. The results are presented as Spearman-$\text{R}_{Z}$ in Table~\ref{tab:text_quality}. Detailed Z-scores for all quality metrics at NTG level are provided in Supplementary Materials.

We observe a strong and statistically significant positive correlation for coherence, grammaticality, as well as SQSE, suggesting that NTGs with lower AA performance are more linguistically and stylistically similar to human writing, thereby complicating NTG-AA. \emph{As sophisticated attackers are likely to employ recent NTGs, their human likeness poses a heightened threat to security systems, potentially enabling evasion of detection and causing misattribution.}

\subsection{Style Distinctiveness}

Style distinctiveness is considered from two perspectives: NTG families and NTG variants.

\subsubsection{Within NTG Families}

\ConfusionMatrix

Many NTGs are developed with minimal changes from their predecessors, often leading to a lack of distinctive styles in new models derived from slightly modified base architectures.
From our AD evaluation, we examined two model families: the proprietary OpenAI GPT-3 model family (including NTGs- davinci-002, Babbage, Curie, text-davinci) and the open-source models derived from fine-tuning the Llama2 model (including NTGs- OpenChat, Vicuna, Stable), which also uses the GPT-3 base architecture. We analyzed the misclassifications within these models to determine if they are contained within the base model. The results are presented in the confusion matrices in Figure~\ref{fig:conf_mat}. The average proportion of misclassifications contained within the base model across all algorithms is 0.43 for GPT-3 and 0.31 for Llama models, indicating greater commonality than distinctiveness within model families.

Although Babbage and Curie are offered as distinct models by OpenAI and treated as such by NTG-AA challenges like AuTexTification, they exhibit the highest rate of misclassifications between themselves, while text-davinci - the newest GPT-3 model, shows nearly equal misclassifications with both Babbage and Curie. A similar trend is observed in the Llama family of models. However, contrary to the expectations, misclassifications among all models within a family are not uniform; notably, older foundational models like davinci-002 and Llama are less frequently predicted. This observation suggests that fine-tuned models or newer iterations of the base architecture do not always retain stylistic commonalities. 

However, \emph{the inconsistent definition of the NTG-AA problem and limited information about NTG evolution—such as base architecture, learning objectives, and finetuning data—hinder style distinctiveness}, complicating reliable NTG attribution.

\subsubsection{Within NTG Parameter Variants}

It is commonly believed that larger models possess enhanced capabilities, potentially leading to better text generation, or in the context of NTG-AA, more refined styles specific to the NTG. If this hypothesis holds, then larger model variants should theoretically be easier to identify. We explore this idea from two perspectives: first, by examining the effect of model size on NTG identification within a base model, and second, by evaluating the similarity between variants of the same base model.
To conduct this analysis, we select seven base models that offer parameter variants across all datasets and utilize the results from WD experiments for our analysis.

\EffectModelVariant

To evaluate the impact of model size on NTG-AA performance, we rank NTG variants by their number of parameters and compute the Spearman-$R$ correlation between each variant’s macro-F1 score and its rank. A strong positive correlation would indicate that larger models, potentially producing more nuanced language, have better identification performance.

Next, to assess the similarities among variants of a base NTG, we calculate two metrics for each base NTG: the proportion of total misclassifications denoted as $\text{Mis}_{\text{T}}$, and the proportion of misclassifications between its variants denoted as $\text{Mis}_{\text{V}}$.
A high $\text{Mis}_{\text{T}}$ suggests that identifying the base NTG is difficult, particularly when each variant is treated as a separate identity. Conversely, a high $\text{Mis}_{\text{V}}$ indicates that samples within the base NTG are often matched to their variant counterparts.
If both these values are high, it suggests greater similarity among NTG variants' language compositions.

The results are presented in Table~\ref{tab:effect_model_variant}.
Contrary to the expectation that larger models would be more easily identified, it is observed that five out of seven base models exhibit a negative correlation, with three being significant. This suggests that smaller models are likely easier to identify. Tying these results with the model sophistication, smaller NTG variants may produce text with lower quality which can help explain why they are comparatively easier to identify.

Furthermore, the values for $\text{Mis}_{\text{T}}$ and $\text{Mis}_{\text{V}}$ in most cases are high and comparable. 
These findings yield dichotomous outcomes, presenting both negative and positive implications. 
\emph{While larger NTG variants may offer attackers more anonymity, the identifiable nature of the base NTG could still provide a means for effective NTG identification}.

\section{Future Research Directions}

Upon thoroughly evaluating the text-based DFIR pipeline under both ideal and CS-ACT attack conditions, we propose several future directions to address existing limitations, particularly in NTD and NTG-AA.

%%%%%%%%%%%%%%%%%%%%%%%%% NTD %%%%%%%%%%%%%%%%%%%%%%%%%
\noindent\textbf{Enhanced Dataset Collection: }
It was observed that the negative impacts of the CS-ACT attack can, to some extent, be mitigated by training the NTD methods on a wide variety of NTGs and diverse data characteristics. While FLAME includes text from 25 distinct NTGs, it lacks diversity in data characteristics. This creates a need for a larger, more comprehensive dataset with data on various topics, domains, and genres generated by past and current NTGs.

\noindent\textbf{Adaptive NTD systems: }
Future work should focus on developing adaptive detection systems that learn from adversarial attacks. This can be done by incorporating adversarial examples in training or by using methods less affected by adversarial perturbations, like content-based approaches. These systems should incorporate nuanced stylistic information combining content, context, structure, and style to improve resilience against adversarial manipulations.

%%%%%%%%%%%%%%%%%%%%%%%%%%%%%%%%%% AA %%%%%%%%%%%%%%%%%%%%%%%%%%%%%%%%%%%
\noindent\textbf{Evaluation on Multi-Lingual Platforms: }
Given that APT targets are global, security systems must be equipped with defenses that operate effectively in a multi-lingual environment. Future research should focus on multi-lingual detection and identification to cover a broader spectrum of cyber threats.

\noindent\textbf{Development of a Knowledge Base for NTG Families: }
The variation in behavior among NTG families, coupled with the lack of detailed information on NTG lineages, underscores the urgent need for a comprehensive knowledge base that systematically documents the family trees of NTGs, whether proprietary or open-source. While firms like OpenAI withhold specific model details, understanding these family trees is crucial for forming robust associations for reliable NTG-AA.

\noindent\textbf{Hierarchical NTG-AA: }
Forensic profiling of NTGs can be improved with a hierarchical approach, first identifying the parent NTG and then the individual child NTGs. This method leverages the greater similarities within NTG families, using a process of elimination to discard similarities rather than finding differences, addressing the challenges of NTG-AA.

\noindent\textbf{Stylizing NTGs: }
Inspired by both computational linguistics and the concept of watermarking, future research could explore the possibility of embedding a distinctive style into LLMs. This style would mimic the way humans acquire language through experiential processes and could be based on factors like personality, age, gender, education, etc. Such an approach would ensure that each model's text generation is uniquely stylistic, enhancing both identification and security measures.

\section{Conclusion}
This paper presents a comprehensive evaluation of the DFIR pipeline for text-based security systems, focusing on the detection of NTG-authored text and the authorship attribution in both human and NTG-authored content. By introducing a novel co-authorship text attack- CS-ACT, that blends human and NTG text, our work highlights that the ideal scenarios, typically assumed by NTD and NTG-AA studies, do not accurately mirror the real-world conditions revealing significant vulnerabilities in current DFIR methodologies.

Our comprehensive evaluation of the DFIR pipeline, employing a robust experimental setup including 22 algorithms implemented on 14 datasets with diverse characteristics and 43 unique NTGs including the latest advancements up to GPT-4o, reveals that while the Incident detection phase remains relatively robust, Forensic profiling phase, specifically NTG-AA, is notably unreliable due to inherent similarities in NTGs. This unreliability is compounded under the CS-ACT attack scenario, providing opportunities for attackers to evade detection and mislead identification efforts. The weakest link in the pipeline, NTG-AA, has been analyzed in depth to uncover the limitations in accurately identifying NTGs. Key factors such as improper formulation of the NTG-AA problem, model sophistication, and the lack of distinctive style in NTGs have been identified as critical issues impacting performance.

This study provides critical insights into the limitations of current NTD and AA algorithms, advancing the field of information forensics and security by underscoring the need for more advanced and adaptive methods. Key recommendations include embedding stylized language in LLMs to enhance traceability, mapping the lineage of LLMs for improved source attribution, and developing adaptive detection methods that learn from new attack vectors.

\bibliographystyle{IEEEtran}
\bibliography{references}

\section{Supplementary Materials}
\subsection{Dataset Imbalance Details}

\begin{table}[!hbp]
  \centering
  \captionsetup[subfloat]{labelfont=scriptsize,textfont=scriptsize,justification=centering}
  \caption{Number of Human and NTG Samples}
  \begin{threeparttable}
    \begin{tabular}{llllll}
    \toprule
    \multicolumn{2}{c}{Dataset}  & \#Texts & Human & NTG \\
    \midrule
    AuTexT \cite{sarvazyan2023overview} & ~ & 55709 & 27688 & 28021 \\
    SynSci \cite{rosati2022synscipass}& ~ & 5889  & 1000 & 4889  \\
    TweepF \cite{fagni2021tweepfake} & ~ & 25572 & 12786  & 12786  \\
    Uchendu \cite{uchendu2020authorship}& ~ & 9594  & 1066  & 8528 \\
    TBench \cite{uchendu2021turingbench}& ~ & 168612 & 159758  & 8854 \\
    \multirow{10}{*}{DFTD \cite{li2023mage}} 
    & CMV  &  21547 & 6449 & 15098  \\
    & ELI5 &  34707 & 16925  & 17782  \\
    & SciGen& 25133 & 6829  & 18304  \\
    & SQuAD&  36420 & 17111  & 19309  \\
    & TLDR&   19481 & 4125  & 15356   \\
    & WP  &   27731 & 9416  & 18315  \\
    & XSum &  27746 & 7698  & 20048   \\
    & Yelp &  43484 & 25868  & 17798  \\
    FLAME & ~ & 59831 & 26245 & 33586 \\
    \bottomrule
    \end{tabular}%
    \end{threeparttable}
  \label{tab:data_stats_imbalance}%
\end{table}%

\subsection{DFTD Preprocessing}

The DFTD dataset, now known as MAGE, followed a generation protocol where the text provided to each NTG for generation was retained by autoregressive and chat models. Each NTG received the same set of human seed text, resulting in NTG text samples with identical seed text across different NTGs. Additionally, the human seed text was included in the ``human'' class, creating an overlap between NTG and human text samples, potentially affecting NTD algorithm performance.

According to the source paper, the first 30 tokens of each text sample were used for NTG text generation. Therefore, these tokens were removed from all samples, and only those with at least 20 tokens remaining, whether human or NTG, were retained.
Originally, the dataset had 10 subsets. However, the HellaSwag and ROC subsets did not contain enough samples and were excluded from this work.

\subsection{FLAME Details}

\noindent\textbf{Detailed Dataset Creation: }

The initial pool of authors was selected from the Blogs Authorship Corpus (BAC). The pre-processing steps included removing duplicate spaces, line breaks, non-ASCII characters, initial punctuations, and specific tokens like ``urlLink''. An \emph{Author Data Repository} was formed for each author where the text samples were retained only if they had a gibberish score\footnote{https://pypi.org/project/gibberish-detector/}$<=$3.00. Next, the text samples were subjected to chunking where samples greater than 400 tokens were broken into independent samples of unique and non-overlapping segments of 400 tokens each. 

Preprocessing was performed to remove duplicate spaces, line breaks, non-ASCII characters, punctuation at the beginning of the samples, and finally BAC-specific token ``urlLink''.
Next, only the authors who had at least 10 text samples exceeding 200 tokens and could provide at least 125 samples with 400 tokens after pre-processing were retained, narrowing down to 135 authors.

For the creation of FLAME$_{Perturb}$, we randomly selected 125 samples of 400 tokens from each author. Each of the 25 NTGs was then tasked with generating text from 5 of these samples, provided with a varying proportion (P) of the original human-authored text as seed. For instance, P=25 meant that H$_{seed}$=100 tokens of human content were used to seed the generation of the next N$_{gen}$=300 tokens by an NTG. This process resulted in five distinct subsets of FLAME$_{Perturb}$, denoted as \emph{FLAME$_{P}$} while allowing a $\pm$10\% buffer in generation length to maintain human-NTG proportionality.

For FLAME$_{Pure}$, we further subdivided the dataset into \emph{FLAME$_{Human}$} and \emph{FLAME$_{NTG}$}. In FLAME$_{NTG}$, 10 samples from each author not used in FLAME$_{Perturb}$ were chosen and the first 50 tokens for each text sample were provided as seed to every NTG. Samples not utilized as prompts in either subset contributed to the FLAME$_{Human}$ set.

\noindent\textbf{Implementation Details: }

As the objective of the study was to generate text by the NTGs as close to the human-authored text as possible, we adopted the author style emulation strategy adopted by \cite{bhandarkar2024emulating}. Following the same protocol, the chat-based models were provided with the instructional prompt ``Your task is to generate a N$_{gen}$-word continuation that seamlessly integrates with the provided snippet. Strive to make the continuation indistinguishable from the human-authored text. As output, only provide the entire completed text without explanation or any other comments.'', where N$_{gen}$ represents the generation length which is determined based on the value of P. This instruction is removed after the text is generated. To ensure consistency in text generation across NTGs, nucleus sampling with top-K (K = 50) and top-p (p = 0.95) is employed as the decoding strategy wherever applicable. 

\begin{table}[t]
\centering
\captionsetup[subfloat]{labelfont=scriptsize,textfont=scriptsize,justification=centering}
\caption{NTGs used in FLAME}
\begin{threeparttable}
\begin{tabular}{p{1.3cm}p{0.65cm}p{5.5cm}}
\toprule
    NTGs & \#Param & Model Identifier \\ 
\midrule
    Bloom & 7B & bigscience/bloom-7b1 \\ 
    BTLM & 3B & cerebras/btlm-3b-8k-base \\ 
    CTRL & 1.63B & Salesforce/ctrl \\ 
    Dolly-v2 & 12B & databricks/dolly-v2-12b \\ 
    Falcon & 7B & tiiuae/falcon-11B \\ 
    Galactica & 6.7B & facebook/galactica-6.7b \\ 
    Gemma & 7B & google/gemma-7b \\ 
    GPT2 & 124M & gpt2 \\ 
    GPT 3.5 & n/a & gpt-3.5-turbo \\ 
    GPT4 & n/a & gpt-4o \\ 
    GPTCerebras & 13B & cerebras/Cerebras-GPT-13B \\ 
    GPT-J & 6B & EleutherAI/gpt-j-6b \\ 
    GPT-NEO & 2.7B & EleutherAI/gpt-neo-2.7B \\ 
    Incite & 7B & togethercomputer/RedPajama-INCITE-7B-Base \\ 
    Llama3 & 13B & meta-llama/Meta-Llama-3-8B \\ 
    Mistral & 7B & mistralai/Mistral-7B-v0.3 \\ 
    MPT & 7B & mosaicml/mpt-7b-8k \\ 
    OPT & 13B & facebook/opt-13b \\ 
    Phi3 & 14B & microsoft/Phi-3-medium-4k-instruct \\ 
    Stable & 12B & stabilityai/stablelm-2-12b \\ 
    XLNET & 340M & xlnet-base-cased \\ 
    CausalLM & 14B & CausalLM/14B \\ 
    OpenChat & 8B & openchat/openchat-3.6-8b \\ 
    Vicuna & 13B & lmsys/vicuna-13b-v1.5 \\ 
    XGLM & 7.5B & facebook/xglm-7.5B \\ 
\bottomrule
\end{tabular}
\end{threeparttable}
\begin{tablenotes}
\item[] Note: Most NTGs are sourced from HuggingFace model repository (https://huggingface.co/models). Proprietary models, GPT3.5 and GPT4-o were sourced from OpenAI (https://platform.openai.com/docs/models).
\end{tablenotes}
\end{table}

\begin{table}[t]
    \centering
    \captionsetup[subfloat]{labelfont=scriptsize,textfont=scriptsize,justification=centering}
    \caption{NTG-level Z-scores with Text Quality Metrics}
    \begin{threeparttable}
    \begin{tabular}{lllllll}
    \toprule
    NTG & Perf\tnote{5} & Coh-W\tnote{1} & Coh-S\tnote{2} & Gram\tnote{3} & Redun\tnote{4} & SQSE \\ 
    \midrule
    XLM & 0.89 & 2.82 & 3.00 & 1.31 & 0.43 & 2.94 \\ 
    SCIgen & 0.85 & 1.18 & 0.87 & 0.58 & 0.50 & 0.78 \\ 
    GPT & 0.83 & 1.56 & 1.81 & 1.27 & 0.27 & 2.10 \\ 
    Others & 0.80 & 1.36 & 1.60 & 2.23 & 0.17 & 2.03 \\ 
    RNN & 0.80 & 2.50 & 2.27 & 2.40 & 0.63 & 1.03 \\ 
    TransXL & 0.78 & 1.82 & 2.90 & 2.28 & 0.17 & 0.61 \\ 
    XGLM & 0.77 & 1.40 & 1.83 & 1.48 & 0.53 & 0.91 \\ 
    PPLM & 0.76 & 1.88 & 2.06 & 2.02 & 0.69 & 1.35 \\ 
    Grover & 0.74 & 1.35 & 1.62 & 1.42 & 0.56 & 1.46 \\ 
    CTRL & 0.74 & 0.94 & 0.91 & 0.83 & 0.48 & 1.17 \\ 
    davinci002 & 0.70 & 0.64 & 0.80 & 0.78 & 0.51 & 0.87 \\ 
    GPT4 & 0.69 & 1.85 & 2.09 & 2.08 & 0.34 & 1.28 \\ 
    XLNet & 0.68 & 1.63 & 2.03 & 1.28 & 0.33 & 2.39 \\ 
    GPTNeoX & 0.64 & 0.46 & 0.66 & 0.75 & 0.19 & 0.58 \\ 
    Phi3 & 0.62 & 0.99 & 1.03 & 0.69 & 0.46 & 1.39 \\ 
    Fair & 0.61 & 1.00 & 1.37 & 1.40 & 0.44 & 0.92 \\ 
    GPT3.5 & 0.51 & 0.73 & 0.56 & 0.88 & 0.39 & 0.62 \\ 
    Galactica & 0.47 & 0.59 & 0.71 & 0.64 & 0.22 & 0.53 \\ 
    OpenChat & 0.46 & 0.67 & 0.72 & 0.88 & 0.32 & 0.66 \\ 
    T0 & 0.39 & 0.68 & 0.70 & 0.62 & 0.23 & 0.54 \\ 
    Falcon & 0.39 & 0.74 & 0.78 & 0.82 & 0.48 & 1.18 \\ 
    FlanT5 & 0.37 & 0.71 & 0.82 & 0.58 & 0.28 & 0.81 \\ 
    GLM & 0.37 & 0.72 & 0.71 & 0.80 & 1.17 & 1.28 \\ 
    Dolly & 0.36 & 0.87 & 0.83 & 0.89 & 0.57 & 1.46 \\ 
    Babbage & 0.35 & 0.99 & 0.59 & 0.76 & 0.52 & 1.14 \\ 
    text-davinci & 0.34 & 1.00 & 0.58 & 0.72 & 0.52 & 1.11 \\ 
    Curie & 0.34 & 0.77 & 0.58 & 0.82 & 0.51 & 0.79 \\ 
    GPT2 & 0.33 & 1.00 & 0.94 & 0.95 & 0.47 & 1.09 \\ 
    Stable & 0.32 & 0.75 & 0.74 & 0.84 & 0.21 & 0.65 \\ 
    Bloom & 0.32 & 0.59 & 0.55 & 0.56 & 0.61 & 1.03 \\ 
    Vicuna & 0.31 & 0.51 & 0.56 & 0.52 & 0.97 & 1.20 \\ 
    GPT-J & 0.30 & 0.68 & 0.71 & 0.81 & 0.18 & 0.56 \\ 
    OPT-IML & 0.26 & 0.57 & 0.55 & 0.63 & 0.58 & 0.88 \\ 
    Llama & 0.22 & 0.69 & 0.77 & 0.77 & 0.87 & 0.90 \\ 
    Mistral & 0.22 & 0.63 & 0.66 & 0.56 & 0.28 & 0.52 \\ 
    OPT & 0.21 & 0.62 & 0.68 & 0.59 & 0.19 & 0.51 \\ 
    Gemma & 0.21 & 0.58 & 0.60 & 0.61 & 0.54 & 0.94 \\ 
    CerebrasGPT & 0.19 & 0.59 & 0.65 & 0.61 & 0.30 & 0.53 \\ 
    CausalLM & 0.16 & 0.73 & 0.73 & 0.67 & 0.27 & 0.58 \\ 
    MPT & 0.15 & 0.63 & 0.72 & 0.61 & 0.32 & 0.54 \\ 
    GPTNeo & 0.14 & 0.64 & 0.62 & 0.61 & 0.31 & 0.55 \\ 
    BLTM & 0.13 & 0.66 & 0.68 & 0.63 & 0.31 & 0.55 \\ 
    Incite & 0.12 & 0.66 & 0.70 & 0.60 & 0.32 & 0.53 \\ 
    \bottomrule
    \end{tabular}
    \begin{tablenotes}
    \item[1]Word Coherence Score; \item[2]Sentence Coherence Score; \item[3]Grammaticality Score; \item[4]Redundancy Score; \item[5] Macro-F1 Performance     
    \end{tablenotes}
\end{threeparttable}
\end{table}

\subsection{Glossary}
\begin{itemize}
    \item AA: Authorship Attribution
    \item AD: Across-dataset
    \item AI: Artificial Intelligence
    \item APT: Advanced Persistent Threat
    \item ASR: Attack Success Rate
    \item CS-ACT: Compromising Security via Adversarially Co-Authored Text
    \item DFIR: Digital Forensics and Incident Response 
    \item FLAME: Forensic LLM-Authorship Mixture Evaluation
    \item GenAI: Generative Artificial Intelligence
    \item LIWC: Linguistic Inquiry Word Count
    \item LLM: Large Language Model
    \item NTG: Neural Text Generator
    \item NTD: Neural Text Detection
    \item TTPs: Tactics, Techniques, and Procedures
    \item WD: Within-dataset
\end{itemize}

\begin{table*}[!ht]
    \centering
    \captionsetup[subfloat]{labelfont=scriptsize,textfont=scriptsize,justification=centering}
    \caption{NTG-level Across-Datasets Performance for NTG-AA. Reported in Macro-F1 score}
    \begin{threeparttable}
    \begin{tabular}{lllllllllllllll}
    \toprule
    AA Methods & LIWC & WRFC & Char-3 & MHC & ContraX & BERT- & BERT- & RoB- & FFLM &  DeB- & XLNet- & T5S & LFI &Avg./\\
    NTG ($\downarrow$)& DT &  & grams & &  & AA & multi & multi & & FT & FT &  &  & NTG\\

    \midrule
    XLM & 0.92 & 0.99 & 0.99 & 0.96 & 1.00 & 1.00 & 1.00 & 1.00 & 0.39 & 1.00 & 1.00 & 0.99 & 0.39 & 0.89 \\ 
    SciGen & 0.91 & 0.97 & 0.94 & 0.99 & 0.98 & 0.99 & 0.99 & 0.99 & 0.39 & 0.98 & 0.99 & 0.98 & 0.00 & 0.85 \\ 
    GPT1 & 0.70 & 0.99 & 0.95 & 0.95 & 0.99 & 0.98 & 0.99 & 1.00 & 0.37 & 0.99 & 1.00 & 0.69 & 0.23 & 0.83 \\ 
    TransfoXL & 0.66 & 0.83 & 0.90 & 0.89 & 0.99 & 0.97 & 0.98 & 0.99 & 0.20 & 0.99 & 0.99 & 0.97 & 0.00 & 0.80 \\ 
    RNN & 0.64 & 0.88 & 0.85 & 0.84 & 0.96 & 0.96 & 0.95 & 0.96 & 0.37 & 0.95 & 0.95 & 0.94 & 0.08 & 0.80 \\ 
    XGLM & 0.54 & 0.74 & 0.83 & 0.83 & 0.99 & 0.97 & 0.97 & 0.98 & 0.38 & 0.99 & 0.99 & 0.94 & 0.00 & 0.78 \\ 
    PPLM & 0.45 & 0.89 & 0.88 & 0.85 & 0.98 & 0.94 & 0.98 & 0.98 & 0.15 & 0.97 & 0.96 & 0.95 & 0.00 & 0.77 \\ 
    Others & 0.75 & 0.86 & 0.52 & 0.39 & 0.96 & 0.96 & 0.95 & 0.95 & 0.38 & 0.95 & 0.94 & 0.76 & 0.54 & 0.76 \\ 
    Grover & 0.38 & 0.83 & 0.83 & 0.61 & 0.97 & 0.95 & 0.94 & 0.93 & 0.28 & 0.94 & 0.89 & 0.95 & 0.18 & 0.74 \\ 
    CTRL & 0.55 & 0.88 & 0.85 & 0.70 & 0.96 & 0.91 & 0.95 & 0.96 & 0.34 & 0.97 & 0.96 & 0.39 & 0.20 & 0.74 \\ 
    GPT4 & 0.47 & 0.93 & 0.71 & 0.63 & 0.87 & 0.79 & 0.88 & 0.90 & 0.33 & 0.86 & 0.88 & 0.91 & 0.01 & 0.70 \\ 
    davinci002 & 0.40 & 0.75 & 0.67 & 0.36 & 0.93 & 0.90 & 0.93 & 0.95 & 0.27 & 0.93 & 0.94 & 0.91 & 0.00 & 0.69 \\ 
    XLNet & 0.46 & 0.77 & 0.68 & 0.72 & 0.88 & 0.83 & 0.87 & 0.88 & 0.13 & 0.88 & 0.89 & 0.80 & 0.07 & 0.68 \\ 
    Phi & 0.38 & 0.75 & 0.66 & 0.47 & 0.84 & 0.69 & 0.90 & 0.89 & 0.17 & 0.86 & 0.89 & 0.85 & 0.00 & 0.64 \\ 
    GPTNeoX & 0.42 & 0.81 & 0.60 & 0.09 & 0.83 & 0.86 & 0.88 & 0.84 & 0.33 & 0.79 & 0.80 & 0.73 & 0.12 & 0.62 \\ 
    Fair & 0.27 & 0.76 & 0.68 & 0.62 & 0.88 & 0.83 & 0.85 & 0.80 & 0.00 & 0.80 & 0.83 & 0.62 & 0.01 & 0.61 \\ 
    GPT3.5 & 0.20 & 0.50 & 0.47 & 0.40 & 0.73 & 0.70 & 0.74 & 0.72 & 0.00 & 0.70 & 0.72 & 0.68 & 0.00 & 0.51 \\ 
    Galactica & 0.08 & 0.46 & 0.45 & 0.40 & 0.69 & 0.49 & 0.68 & 0.75 & 0.08 & 0.70 & 0.75 & 0.53 & 0.05 & 0.47 \\ 
    OpenChat & 0.16 & 0.48 & 0.45 & 0.42 & 0.52 & 0.45 & 0.74 & 0.78 & 0.24 & 0.61 & 0.70 & 0.47 & 0.00 & 0.46 \\ 
    Falcon & 0.13 & 0.30 & 0.43 & 0.41 & 0.56 & 0.35 & 0.65 & 0.67 & 0.19 & 0.66 & 0.66 & 0.07 & 0.00 & 0.39 \\ 
    T0 & 0.16 & 0.34 & 0.18 & 0.07 & 0.67 & 0.59 & 0.58 & 0.59 & 0.09 & 0.65 & 0.64 & 0.47 & 0.04 & 0.39 \\ 
    Dolly & 0.09 & 0.20 & 0.26 & 0.27 & 0.69 & 0.50 & 0.57 & 0.71 & 0.21 & 0.57 & 0.68 & 0.08 & 0.00 & 0.37 \\ 
    GLM & 0.12 & 0.55 & 0.26 & 0.26 & 0.47 & 0.40 & 0.51 & 0.56 & 0.17 & 0.47 & 0.43 & 0.43 & 0.20 & 0.37 \\ 
    Flan & 0.16 & 0.46 & 0.16 & 0.02 & 0.68 & 0.60 & 0.57 & 0.56 & 0.13 & 0.62 & 0.59 & 0.15 & 0.00 & 0.36 \\ 
    Babbage & 0.23 & 0.43 & 0.29 & 0.13 & 0.53 & 0.51 & 0.50 & 0.42 & 0.10 & 0.44 & 0.53 & 0.40 & 0.00 & 0.35 \\ 
    Curie & 0.22 & 0.45 & 0.24 & 0.19 & 0.49 & 0.49 & 0.38 & 0.43 & 0.12 & 0.51 & 0.41 & 0.46 & 0.00 & 0.34 \\ 
    text-davinci & 0.13 & 0.30 & 0.24 & 0.10 & 0.55 & 0.51 & 0.50 & 0.48 & 0.15 & 0.50 & 0.47 & 0.43 & 0.02 & 0.34 \\ 
    GPT2 & 0.12 & 0.18 & 0.22 & 0.13 & 0.66 & 0.54 & 0.58 & 0.60 & 0.00 & 0.59 & 0.62 & 0.02 & 0.00 & 0.33 \\ 
    Stable & 0.07 & 0.29 & 0.37 & 0.23 & 0.50 & 0.35 & 0.56 & 0.56 & 0.00 & 0.58 & 0.56 & 0.11 & 0.00 & 0.32 \\ 
    Bloom & 0.19 & 0.46 & 0.28 & 0.22 & 0.50 & 0.40 & 0.47 & 0.39 & 0.00 & 0.28 & 0.49 & 0.46 & 0.01 & 0.32 \\ 
    GPT-J & 0.26 & 0.51 & 0.37 & 0.34 & 0.31 & 0.39 & 0.40 & 0.35 & 0.17 & 0.30 & 0.32 & 0.21 & 0.07 & 0.31 \\ 
    Vicuna & 0.08 & 0.37 & 0.27 & 0.17 & 0.48 & 0.28 & 0.63 & 0.59 & 0.10 & 0.47 & 0.48 & 0.03 & 0.00 & 0.30 \\ 
    OPT-IML & 0.14 & 0.36 & 0.20 & 0.04 & 0.45 & 0.38 & 0.46 & 0.43 & 0.09 & 0.42 & 0.41 & 0.02 & 0.03 & 0.26 \\ 
    Llama & 0.09 & 0.41 & 0.20 & 0.20 & 0.26 & 0.38 & 0.28 & 0.25 & 0.09 & 0.18 & 0.24 & 0.26 & 0.06 & 0.22 \\ 
    Gemma & 0.08 & 0.26 & 0.22 & 0.23 & 0.28 & 0.22 & 0.31 & 0.35 & 0.04 & 0.29 & 0.32 & 0.24 & 0.00 & 0.22 \\ 
    Mistral & 0.09 & 0.25 & 0.11 & 0.04 & 0.23 & 0.16 & 0.44 & 0.46 & 0.08 & 0.30 & 0.33 & 0.22 & 0.00 & 0.21 \\ 
    OPT & 0.11 & 0.25 & 0.15 & 0.05 & 0.30 & 0.31 & 0.28 & 0.25 & 0.00 & 0.26 & 0.31 & 0.35 & 0.05 & 0.21 \\ 
    CerebrasGPT & 0.08 & 0.17 & 0.18 & 0.15 & 0.27 & 0.17 & 0.25 & 0.30 & 0.12 & 0.29 & 0.30 & 0.14 & 0.00 & 0.19 \\ 
    CausalLM & 0.06 & 0.09 & 0.15 & 0.12 & 0.24 & 0.17 & 0.26 & 0.28 & 0.07 & 0.21 & 0.22 & 0.18 & 0.00 & 0.16 \\ 
    MPT & 0.07 & 0.13 & 0.16 & 0.14 & 0.18 & 0.13 & 0.26 & 0.29 & 0.05 & 0.27 & 0.29 & 0.00 & 0.00 & 0.15 \\ 
    GPT-Neo & 0.07 & 0.19 & 0.14 & 0.15 & 0.21 & 0.14 & 0.22 & 0.15 & 0.08 & 0.18 & 0.18 & 0.14 & 0.00 & 0.14 \\ 
    BTLM & 0.06 & 0.12 & 0.09 & 0.10 & 0.17 & 0.12 & 0.18 & 0.25 & 0.12 & 0.15 & 0.15 & 0.15 & 0.00 & 0.13 \\ 
    Incite & 0.08 & 0.15 & 0.09 & 0.12 & 0.17 & 0.12 & 0.11 & 0.17 & 0.10 & 0.09 & 0.17 & 0.20 & 0.00 & 0.12 \\ 
    \bottomrule
    \end{tabular}
\end{threeparttable}
\end{table*}

\end{document}